\documentclass[aps, twocolumn, prd, preprintnumbers, amsmath, amssymb, amsfonts, superscriptaddress, nofootinbib]{revtex4-2}
\pdfoutput=1

\usepackage[pagebackref=false,hidelinks]{hyperref}

\usepackage{setspace,latexsym}
\usepackage{color}
\usepackage{epsfig}
\usepackage{graphicx}
\usepackage{slashed}
\usepackage[export]{adjustbox}
\usepackage{soul}

\newcommand{\beq}{\begin{equation}}
\newcommand{\eeq}{\end{equation}}
\newcommand{\bea}{\begin{eqnarray}}
\newcommand{\eea}{\end{eqnarray}}

\newcommand{\GeV}{\mathrm{GeV}}

\newcommand{\C}{\mathrm{c}}
\newcommand{\I}{\mathrm{i}}
\newcommand{\e}{\mathrm{e}}
\newcommand{\dd}{\mathrm{d}}

\newcommand{\eff}{\mathrm{eff}}
\newcommand{\Mpl}{M_\mathrm{Pl}}

\newcommand{\DM}{\mathrm{DM}}
\newcommand{\PBH}{\mathrm{PBH}}
\newcommand{\Tstarmax}{T_{*\mathrm{max}}}

\newcommand{\lsim}{\lesssim}
\newcommand{\gsim}{\gtrsim}

\renewcommand{\matrix}[1]{\!\begin{pmatrix} #1 \end{pmatrix}\!}

\def\bal#1\eal{\begin{align}#1\end{align}}

\begin{document} 

\title{Primordial black holes from bubble collisions during a first-order phase transition}

\author{Tae Hyun Jung}
\email{thjung0720@gmail.com}
\affiliation{Department of Physics, Florida State University, Tallahassee, FL 32306, USA}
\affiliation{Center for Theoretical Physics of the Universe, Institute for Basic Science (IBS), Daejeon, 34126, Korea}

\author{Takemichi Okui}
\email{tokui@fsu.edu}
\affiliation{Department of Physics, Florida State University, Tallahassee, FL 32306, USA}
\affiliation{Theory Center, High Energy Accelerator Research Organization (KEK), Tsukuba 305-0801, Japan}

\preprint{KEK-TH-2350}

\begin{abstract}
We study the possibility of production of primordial black holes (PBHs) from bubble collisions during a first-order phase transition. While typical colliding bubbles are small and irrelevant for PBH production, we find that those that can produce PBHs must have a macroscopically thick fluid shell and have been born much before the typical nucleation time. Particularly large uncertainties arise from an exponential sensitivity of the nucleation rate on the required duration of bubble growth which depends on the details of the collisions and the evolution of the spacetime metric toward the end of the phase transition. We introduce a few parameters to be obtained from future numerical simulation to represent those unknowns, and estimate the PBH abundance in an Abelian Higgs benchmark model and show that it can be significant. We predict an approximately monochromatic PBH mass spectrum, and find regions in the parameter space where the PBHs can constitute entire dark matter or even over-close the universe. Our result thus shows that models with a first-order phase transition can be constrained by over-abundant PBHs or null results of other PBH searches. 
\end{abstract}

\maketitle

%%%%%%%%%%%%%%%%%%%%%%%%%%%%%%%%%%%%%%%%%%%%%%%%%%%%%%%%%%%%%%%%%%%%%%%%
%%%%%%%%%%%%%%%%%%%%%%%%%%%%%%%%%%%%%%%%%%%%%%%%%%%%%%%%%%%%%%%%%%%%%%%%
\section{Introduction}
First-order phase transition in the early universe is a well motivated field of study both theoretically and experimentally.
Since a generic quantum field theory has multiple phases, it is natural to expect the early universe to undergo various stages of phase transitions, 
some of which may well be at first order.
A first-order phase transition necessarily leads to nucleation of true-vacuum bubbles and their subsequent expansions, and they must eventually collide to fill the universe with the true vacuum.
This can result in many cosmological implications such as 
dark matter~\cite{Chway:2019kft, Baker:2019ndr, Ahmadvand:2021vxs, Hong:2020est, Azatov:2021ifm,Asadi:2021pwo} and
baryogenesis~\cite{Shaposhnikov:1987tw, Cline:2018fuq, Hall:2019ank, Fujikura:2021abj, Arakawa:2021wgz, Azatov:2021irb}.
Since gravitational wave production from bubble collisions during a first-order phase transition is generically expected~\cite{Witten:1984rs,Hogan:1986qda,Kosowsky:1991ua,Kosowsky:1992vn,Kamionkowski:1993fg,Espinosa:2010hh,Caprini:2019egz, Schmitz:2020syl},
these ideas may be testable by near-future gravitational wave experiments~\cite{%
Harry:2010zz, LIGOScientific:2014pky,VIRGO:2014yos,%Advanced LIGO/Virgo
LISA:2017pwj, Baker:2019nia,%LISA
Seto:2001qf, Kawamura:2011zz,Yagi:2011wg,Isoyama:2018rjb,%DECIGO
Crowder:2005nr, Corbin:2005ny,Harry:2006fi,%BBO
Somiya:2011np, Aso:2013eba,KAGRA:2018plz,KAGRA:2019htd,Michimura:2019cvl,%KAGRA
LIGOScientific:2016wof, Reitze:2019iox,%Cosmic explorer
Punturo:2010zz, Hild:2010id,Sathyaprakash:2012jk,Maggiore:2019uih,%ET
Burke-Spolaor:2018bvk,%PTA
McLaughlin:2013ira, NANOGRAV:2018hou,Aggarwal:2018mgp,Brazier:2019mmu,%NANOGrav
Manchester:2012za, Shannon:2015ect,%Parkes
Kramer:2013kea, Lentati:2015qwp,Babak:2015lua,%EPTA
Hobbs:2009yy, Manchester:2013ndt,Verbiest:2016vem,Hazboun:2018wpv,%IPTA
Carilli:2004nx, Janssen:2014dka,Weltman:2018zrl%SKA
}
(see \cite{Caprini:2019egz, Schmitz:2020syl} for reviews).

Since every first-order phase transition must end with bubble collisions,
it is important to seek other significant outcomes of bubble collisions. 
In particular, given many PBH bounds and searches~\cite{
Carr:2020gox, Carr:2020xqk,%Review
Marani:1998sh, Nemiroff:2001bp, Barnacka:2012bm, Katz:2018zrn, Griest:2013esa, Griest:2013aaa, Niikura:2017zjd, Paczynski:1985jf, Hamadache:2006fw,Wyrzykowski:2009ep, CalchiNovati:2009kq,Wyrzykowski:2010bh,Wyrzykowski:2010mh, Wyrzykowski:2011tr, Niikura:2019kqi, Zumalacarregui:2017qqd, Garcia-Bellido:2017imq,%Lensing
Page:1976wx, Carr:2009jm, Carr:2016hva, Boudaud:2018hqb, DeRocco:2019fjq, Laha:2019ssq, Dasgupta:2019cae, Laha:2020ivk, Chan:2020zry, Belotsky:2014twa, Kim:2020ngi%Evaporation
},
we would like to know whether or not bubble collisions can produce an observationally significant abundance of PBHs, a subject initiated by~\cite{Hawking:1982ga}.
One might be tempted to disregard this possibility and favor other PBH formation mechanisms~\cite{
Sato:1981bf,Maeda:1981gw, Kodama:1982sf,Hall:1989hr,Kusenko:2020pcg}
\cite{Khlopov:1999ys,Lewicki:2019gmv} 
\cite{Gross:2021qgx, Baker:2021nyl,Kawana:2021tde,Baker:2021sno} 
\cite{Liu:2021svg, Hashino:2021qoq, Huang:2022him, DeLuca:2022bjs, Kawana:2022olo, Lewicki:2023ioy, Gouttenoire:2023naa, Jinno:2023vnr} 
as it seems highly improbable for a large amount of energy to be compressed into a small volume by bubble collisions.
In fact, the number of PBHs per Hubble volume \emph{at the time of their formation} must be extremely small:
\beq
\frac{n_\PBH}{H^3} \sim
10^{-15}  f_\PBH 
\!\left( \frac{10^{-11}M_{\odot}}{M_\PBH}\right)\!
\!\left(\frac{10^5\, \rm GeV}{T_*}\right)^{\!\!3}\!
\!\left(\frac{100}{g_*}\right)^{\!\!\frac12} \!\!,
\label{eq:nPBH_need}
\eeq
where $f_\PBH \equiv \rho_\mathrm{PBH}/\rho_\mathrm{DM}$,
and $H$, $T_*$ and $g_*$ are respectively the Hubble rate at the formation time, the temperature and radiation degrees of freedom at the reheating after the phase transition.
Here, we have assumed that the reheating is instantaneous and 100\% efficient, and the PBH number in the comoving volume is constant after that, i.e., $n_\PBH/s = {\rm const}$, ignoring the evaporation effect (which can be neglected only for $M_{\rm PBH} > 10^{-17} M_{\odot}$\,\cite{Carr:2009jm, Carr:2016hva}).
Note that, to avoid the overclosure of the Universe, we need $f_{\rm PBH} \leq 1$, which provides an extremely small upper bound on $H^{-3} n_\PBH$.
For instance, when $M_{\rm PBH}=10^{-11}M_\odot$, $T_*=10^5\,\GeV$ and $g_*=100$, the probability of a successful PBH formation per one Hubble patch at the formation time must be smaller than $10^{-15}$.
Therefore, the high improbableness may be exactly what we need to explain this required rarity of PBHs.
The rarity means a careful estimation is necessary,
and in this Letter we wish to provide such an estimate and identify necessary conditions for an observationally significant PBH production from bubble collisions.

%%%%%%%%%%%%%%%%%%%%%%%%%%%%%%%%%%%%%%%%%%%%%%%%%%%%%%%%%%%%%%%%%%%%%%%%
%%%%%%%%%%%%%%%%%%%%%%%%%%%%%%%%%%%%%%%%%%%%%%%%%%%%%%%%%%%%%%%%%%%%%%%%
\section{The hoop conjecture and its implications}

\subsection{The hoop conjecture}
As a criterion for black hole formation, we employ \emph{the hoop conjecture}~\cite{Thorne:1972ji} ---
a black hole will form in a dense region ${\cal D}$ if 
${\cal D}$ contains mass $M$ satisfying   
$2\pi \cdot 2G M \gtrsim C_{\cal D}$, where $C_{\cal D}$ is the largest circumference of ${\cal D}$.
The whole utility of this criterion relies on \emph{guesstimating} ${\cal D}$, $C_{\cal D}$, and $M$
without having precise knowledge of $g_{\mu\nu}$ or $T_{\mu\nu}$,
because if we knew $g_{\mu\nu}$ or $T_{\mu\nu}$ precisely we would already know whether or not there is a black hole in the spacetime.
In particular, we do not need to consider the curvature of spacetime due to a black hole that might form in ${\cal D}$,
because the inequality in the hoop criterion is an expression of \emph{inconsistency} in the assumption of the \emph{absence} of a black hole.
In~\cite{East:2012mb}, expectations from the hoop conjecture are shown to agree with numerical simulation in the simple case of the collision of two ultrarelativistic balls of perfect fluid.

%%%%%%%%%%%%%%%%%%%%%%%%%%%%%%%%%%%%%%%%%%%%%%%%%%%%%%%%%%%%%%%%%%%%%%%%
%%%%%%%%%%%%%%%%%%%%%%%%%%%%%%%%%%%%%%%%%%%%%%%%%%%%%%%%%%%%%%%%%%%%%%%%
\subsection{Necessity for large bubbles with a thick fluid shell}
Adopting the hoop conjecture, we would like to argue that in order to have any chance of producing a PBH, colliding bubbles must each have a ``large'' radius and a thick shell of plasma on the bubble wall.
(We will define ``large'' shortly.)

\begin{figure}[t] 
\begin{center}
\includegraphics[width=0.45\textwidth]{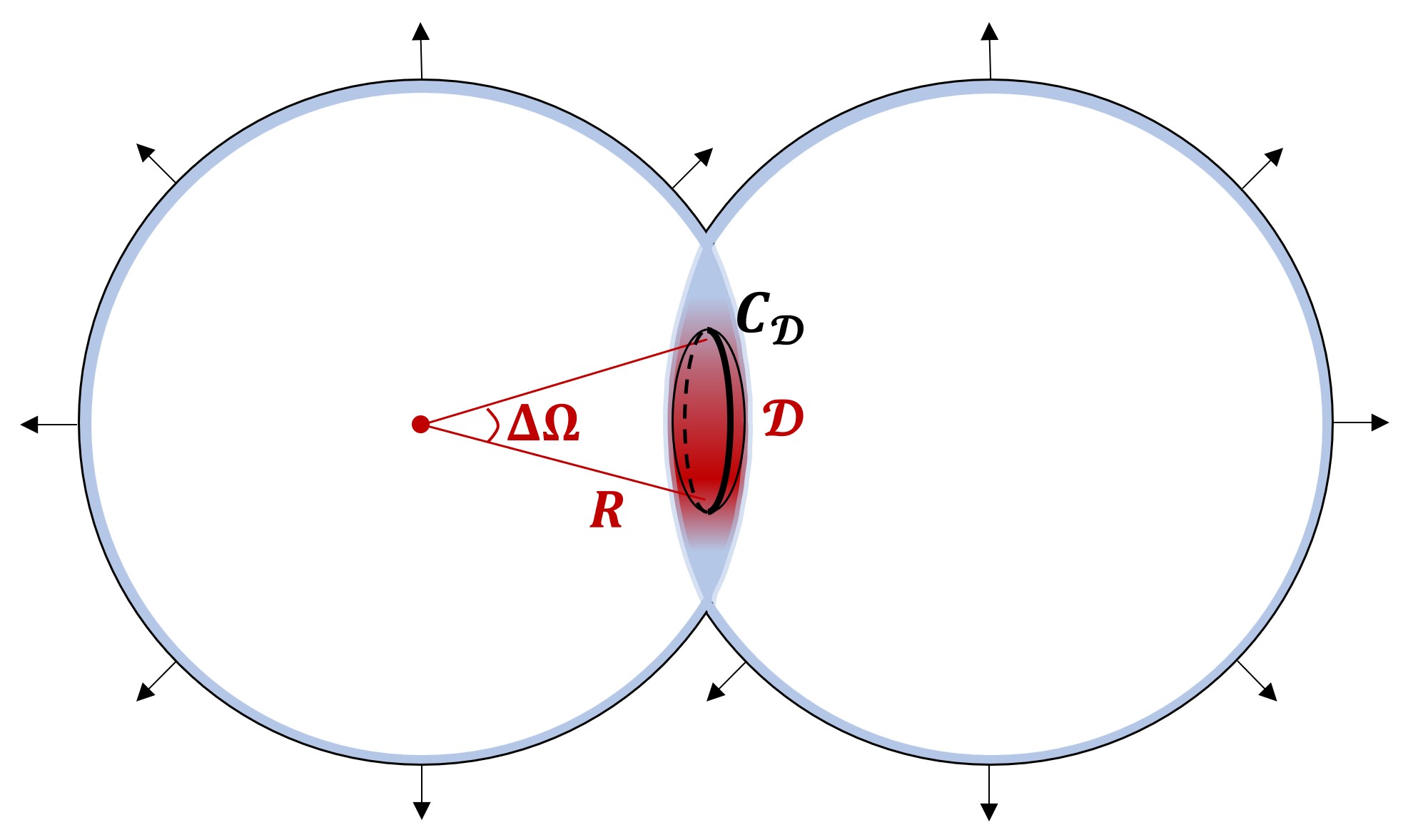} 
\end{center}
\caption{Collision of two bubbles. The bubble walls come with thick fluid shells indicated by light blue.
The red region indicated by ${\cal D}$ describes the energetically dense region that will collapse and form a PBH\@. 
$C_{\cal D}$ denotes the maximum circumference of ${\cal D}$, 
while $R$ and $\Delta\Omega$ denote the distance to and solid angle of ${\cal D}$ from the center of one of the bubbles.}
\label{two_bubbles_cartoon}
\end{figure}

Consider the collision of two bubbles as shown schematically in Fig.~\ref{two_bubbles_cartoon}. 
After the two bubbles collided, part of the region that experienced the collision may form a \emph{dense} region ${\cal D}$ as marked in red.
Let $M$ and $C_{\cal D}$ be the mass and largest circumference of this dense region,
while $\Delta\Omega$ be the solid angle of ${\cal D}$ seen from the center of one of the bubbles.
For simplicity the two radii are assumed to be equal, but this is not essential for
our argument here. Our detailed analysis in Sec.~\ref{sec:PBHnumberdensity} will not use this simplification.

It should be already clear here that \emph{the formation of ${\cal D}$ with any significant $M$ is possible only if the bubble walls come with a macroscopically thick fluid shell} (depicted in light blue in Fig.~\ref{two_bubbles_cartoon}.)
If the bubble walls were made only of the scalar field underlying the phase transition, they would only have a microscopic thickness and hence would only give a negligible $M$.
In a flat spacetime background, 
it is known that a macroscopically thick fluid shell can be formed if there is a large friction that increases linearly or faster with the $\gamma$ factor of the wall velocity~\cite{Kamionkowski:1993fg, Espinosa:2010hh}. 
This is a very nontrivial property and hence restricts possible models (see, e.g., Ref.~\cite{Ai:2023suz} for the case where it is only a logarithmic increase).
It is known in the flat spacetime case that
such property can be realized if the phase transition spontaneously breaks gauge symmetry~\cite{Bodeker:2017cim, Hoche:2020ysm,  Gouttenoire:2021kjv, Azatov:2023xem}.
For large bubbles relevant to our scenario, however, it is important to re-investigate this problem in a curved background, which we leave for future work.

We now demonstrate the necessity of large bubbles by arguing that if the bubbles' radii are small and well within $H^{-1}$ the hoop criterion would lead to contradiction. 
If the bubbles are small, we could assume a flat spatial geometry and energy conservation to guesstimate $M$ and $C_{\cal D}$.
Then, $M$ would be some fraction $\kappa$ of the total energy of the false vacuum that had ever been contained within the red cone given by $C_{\cal D}$ and $\Delta\Omega$, multiplied by two.
Thus, using Euclidean geometry, we would have
\bea
M \sim 2\kappa \cdot \frac{4\pi R^3}{3} \Delta V \!\cdot\! \frac{\Delta \Omega}{4\pi}
\,,\label{M_naive}
\eea
where $R$ is the (slant) height of the red cone as depicted in Fig.~\ref{two_bubbles_cartoon}, and $\Delta V$ is the energy density of the false vacuum, taking that of the true vacuum to be zero.
We could also guesstimate that $C_{\cal D} \simeq 2\pi r$ with $\pi r^2 \simeq R^2\Delta\Omega$.
Then, the application of the hoop criterion to this situation would give 
\bea
1 \lsim
\frac{2\pi\cdot2GM}{C_{\cal D}} 
\sim
\kappa H^2 R^2 \sqrt{\frac{\Delta\Omega}{4\pi}} 
\,\label{eq:criterion_naive}
\eea
with $H^2 \sim 8\pi G \Delta V/3$.
But since $\kappa < 1$ and $\Delta \Omega < 4\pi$, 
this inequality is severely violated for $R \ll H^{-1}$.

Therefore, we must have $R \gsim\! H^{-1}$ to have any chance of satisfying~(\ref{eq:criterion_naive}).%
\footnote{It is not possible to satisfy the criterion by having $N$ bubbles collide with $N \gg 1$ 
as the probability for such an event is given by an exponentially small bubble nucleation rate raised to the $N$-th power multiplied by a tiny probability for $N$ bubbles to collide at a point.
An optimistic conclusion of \cite{Hawking:1982ga} was based on picking up an $O(1)$ fraction of all energy of all $N$ colliding bubbles.}
For $R \gtrsim\! H^{-1}$, we must carefully take into account the curvature of the spacetime and define what we mean by $R$.
We should also pay attention to causal relations between the two bubbles.

The necessity for $R \gtrsim\! H^{-1}$ has important physical implications for the nature of the phase transition.
On the one hand, in order for such large bubbles to have grown so much by eating up the false vacuum energy, the universe must have been in the false vacuum for a long time.
On the other hand, for a successful completion of the phase transition without eternal inflation, the universe must be filled in by many small bubbles quickly.
The latter means that the nucleation time $t_{\rm n}$ (i.e., when one or more bubbles begin to pop up in every horizon) must not be too different from the end time $t_{\rm f}$ of the transition, i.e., $t_{\rm f} - t_{\rm n} \lsim O(H^{-1})$.
The former then requires that those large bubbles must have been growing for a long time before $t_{\rm n}$, so the universe must have been in the false vacuum, or in a supercooled phase, for a long duration before $t_{\rm n}$.

To summarize, PBH formation from bubble collisions require large bubbles with a thick fluid shell. 
This means that the scalar field responsible for the phase transition must possess the following three properties. 
First, it must have sufficient couplings to other particles to generate a thick fluid shell on bubble walls.
Second, it must give rise to a long supercooled phase upon the phase transition.
Third, the potential barrier between the true and false vacua must decrease sufficiently rapidly as the temperature drops after $t=t_{\rm n}$ to ensure a successful percolation of the true vacuum without eternal inflation.

%%%%%%%%%%%%%%%%%%%%%%%%%%%%%%%%%%%%%%%%%%%%%%%%%%%%%%%%%%%%%%%%%%%%%%%%
%%%%%%%%%%%%%%%%%%%%%%%%%%%%%%%%%%%%%%%%%%%%%%%%%%%%%%%%%%%%%%%%%%%%%%%%
\subsection{Effects of spacetime curvature for large bubbles}
Here we discuss how the naive expressions~(\ref{M_naive}) and~(\ref{eq:criterion_naive}) should be modified for large bubbles to take into account the spacetime curvature.

\begin{figure}[t] 
\begin{center}
\includegraphics[width=0.4\textwidth]{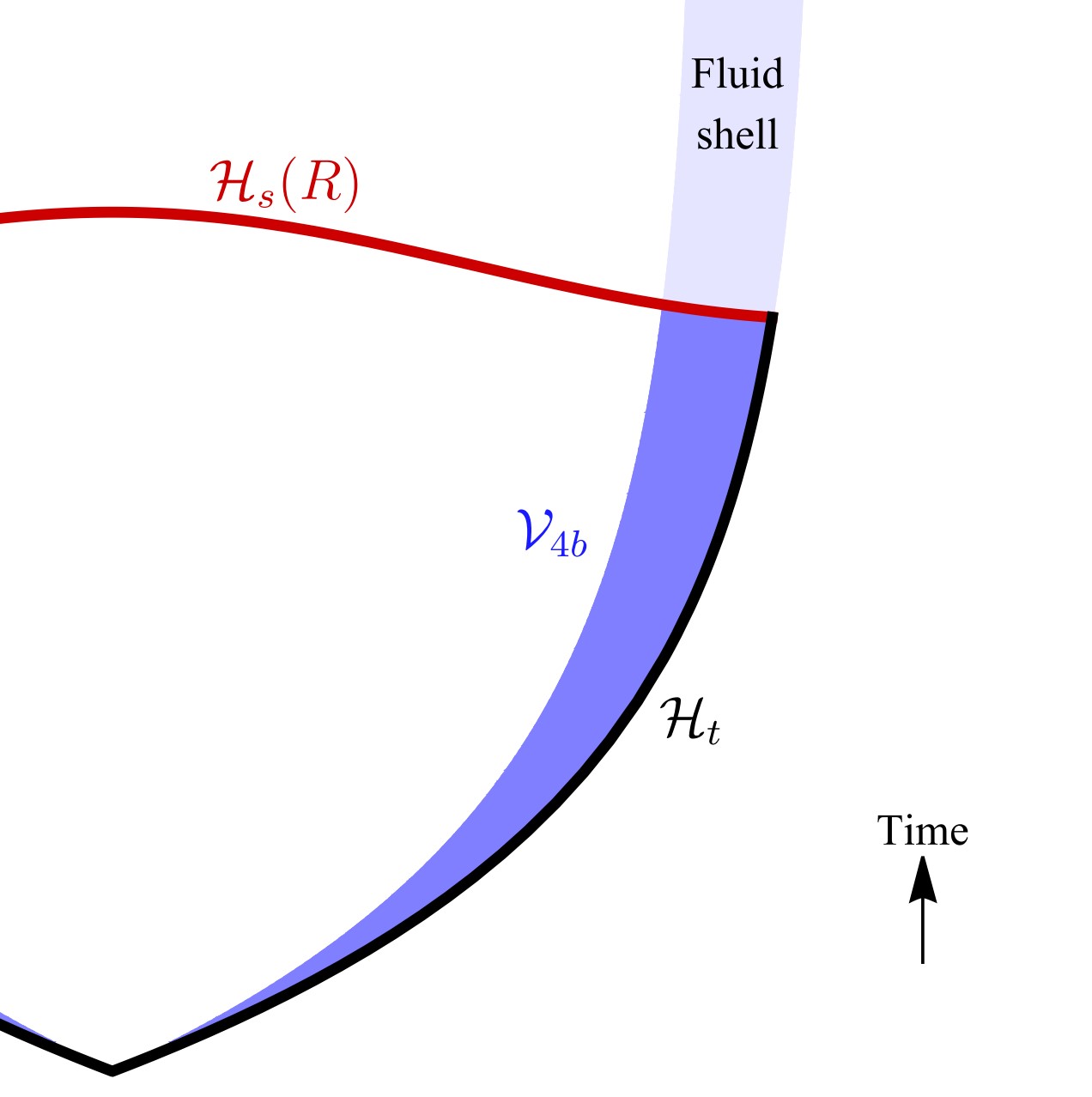} 
\end{center}
\caption{Schematic picture of a single bubble expansion. ${\cal H}_{\rm s}(R)$ is a spacelike hypersurface on which the bubble appears spherically symmetric with radius $R$, ${\cal H}_{\rm t}$ is the hypersurface along the outer surface of the shell, and ${\cal V}_{\rm 4b}$ is the world volume of the fluid shell bounded by $\mathcal{H}_{\rm s}(R)$.}
\label{fig:volumes}
\end{figure}

Fig.~\ref{fig:volumes} schematically shows the structure of one of the colliding bubbles before the collision,
where $\mathcal{H}_{\rm t}$ is the 3d hypersurface given by the trajectory of the outer surface of the shell,
and $\mathcal{H}_{\rm s}(R)$ is a spacelike 3d hypersurface on which the bubble appears spherically symmetric with radius $R$, where $R$ is defined such that the area of the intersection of $\mathcal{H}_{\rm s}(R)$ and $\mathcal{H}_{\rm t}$ is given by $4\pi R^2$.
Finally, $\mathcal{V}_{\rm 4b}$ denotes the 4d world-volume of the bubble shell bounded by $\mathcal{H}_{\rm s}(R)$.

Then, in~(\ref{M_naive}), the naive factor of $4\pi R^3 \Delta V / 3$ representing the total energy of a single bubble with radius $R$ should be replaced as
\bea
\frac{4\pi R^3}{3} \Delta V 
\>\longrightarrow\>
E(R) \equiv \int_{\mathcal{H}_{\rm s}(R)} \hspace{-3.5ex} \dd^3 x \sqrt{\gamma^{\rm (s)}} \; T^{\mu\nu} N_\mu N_\nu
\,,\label{eq:originalE(R)}
\eea
where $T^{\mu\nu}$ is the energy-momentum tensor, $\gamma_{\mu\nu}^{\rm (s)}$ denotes the 3d metric on $\mathcal{H}_{\rm s}(R)$, 
and $N_\mu$ is a normalized timelike 4-vector field orthogonal to $\mathcal{H}_{\rm s}(R)$.

To calculate $E(R)$, we make use of the Gauss theorem on $\nabla_{\!\mu} (T^{\mu\nu} v_\nu)$, 
where $v^\mu$ is a timelike 4-velocity field with the boundary condition that $v^\mu = N^\mu$ on $\mathcal{H}_{\rm s}(R)$. (The boundary condition of $v^\mu$ at $\mathcal{H}_{\rm t}$ will be specified later.)
Then, $E(R)$ can be rewritten as
\beq
E(R)
= \!\int_{\mathcal{H}_{\rm t}} \hspace{-1.5ex} \dd^3x \sqrt{-\gamma^{\rm (t)}} \, n_\mu T^{\mu\nu} v_\nu
  +\!\int_{\mathcal{V}_{\rm 4b}} \hspace{-2.2ex} \dd^4 x \sqrt{-g} \, \nabla_{\!\mu} (T^{\mu\nu} v_\nu),
\label{eq:E(R)_to-be-calculated}  
\eeq
where $n_\mu$ is a normalized spacelike 4-vector field orthogonal to $\mathcal{H}_{\rm t}$ and pointing inward into the bubble.  
We have set $T^{\mu\nu}=0$ in the interior of the bubble (i.e., between $\mathcal{H}_{\rm s}(R)$ and the inner surface of the shell ${\cal V}_{\rm 4b}$), where the spacetime is in the true vacuum.

While we cannot calculate the $\mathcal{V}_{\rm 4b}$ term in~(\ref{eq:E(R)_to-be-calculated}) without a numerical simulation of the bubble shell, 
the $\mathcal{H}_{\rm t}$ term can be analytically evaluated as follows.
Since the exterior of bubbles must be in a highly supercooled phase as we discussed earlier,
we can approximate $T^{\mu\nu}$ outside $\mathcal{H}_{\rm t}$ as $T^{\mu\nu} = -\Delta V g^{\mu\nu}$.
Then, the metric outside $\mathcal{H}_{\rm t}$ can be approximated as de Sitter:
\bea 
\dd s^2 = -\dd t^2 + \e^{2Ht} \bigl[ \dd r^2 + r^2 (\dd\theta^2 + \sin^{2\!}\theta \, \dd\phi^2) \bigr]
\label{eq:deSitter}
\eea
with $H^2 = 8\pi G \Delta V / 3$,
where, without loss of generality thanks to the symmetry of the de Sitter spacetime, 
we have chosen the coordinate system in which the bubble appears spherically symmetric.
We take $v^\mu$ on and outside $\mathcal{H}_{\rm t}$ to be at rest in \emph{this} coordinate system.
While the value of $E(R)$ evidently depends on the choice of $v^\mu$, our choice is arguably the most natural one.
With this choice and in the de Sitter approximation, a straightforward calculation in Appendix~\ref{app:E(R)} gives
\bea
\int_{\mathcal{H}_{\rm t}} \hspace{-1.5ex} \dd^3x \sqrt{-\gamma} \, n_\mu T^{\mu\nu} v_\nu
\>=\> -4\pi \Delta V \!\int_0^{r_0} \hspace{-1ex} \dd r \; r^2 \e^{3Ht(r)}
\,,\label{eq:V3b}
\eea
where $t(r)$ describes $\mathcal{H}_{\rm t}$ as a function of $r$,
and $r_0$ is given via $R = r_0 \e^{Ht(r_0)}$.
The strongly supercooled universe permits us to approximate that the bubble expands at the speed of light, i.e.,
$\dd r = \e^{-Ht} \dd t$ on $\mathcal{H}_{\rm t}$,
which gives $\e^{Ht(r)} = 1/(1 - Hr)$.   
With this approximation, the right-hand side of~(\ref{eq:V3b}) becomes
\bea
-4\pi \Delta V \frac{1}{H^3} \!\left[ \frac{H^2 R^2}{2} - HR + \ln(1+HR) \right] .
\eea
This is \emph{negative} for all values of $HR$,
and for large $HR \gg 1$ (which will be well satisfied as we will see shortly) it can be approximated by
\bea
-\frac{4\pi R^3}{3} \Delta V \frac{3}{2HR}
\,.\label{eq:negative}
\eea

We can now estimate $E(R)$ as follows.
Since the $\mathcal{H}_{\rm t}$ contribution in~(\ref{eq:E(R)_to-be-calculated}) is negative, 
the $\mathcal{V}_{\rm 4b}$ contribution must be positive with its order of magnitude being at least that of the $\mathcal{H}_{\rm t}$ contribution
in order for $E(R)$ to be positive.
We can thus use the magnitude of~(\ref{eq:negative}) as our estimate of $E(R)$ for $HR \gg 1$.
Since we want conditions for PBH \emph{formation,} 
this estimate will lead to \emph{conservative} conditions 
unless there is a fine cancellation between the positive and negative contributions,
which seems extremely unlikely because the $\mathcal{V}_{\rm 4b}$ contribution would depend on the details of the bubble structure and dynamics while the $\mathcal{H}_{\rm t}$ contribution does not, as our above calculation shows.
Therefore, for $HR \gg 1$, our order-of-magnitude estimate is
\bea
E(R) \sim \frac{4\pi R^3}{3} \Delta V \frac{1}{HR}
\,.\label{eq:E(R)}
\eea

Finally, applying the replacement~\eqref{eq:originalE(R)} to~\eqref{M_naive} 
and using the above estimate for $E(R)$, 
we find that the mass estimate~(\ref{M_naive}) should be modified as
\bea
M \sim 2\kappa \cdot \frac{4\pi R^3}{3} \Delta V \frac{1}{HR} \cdot \frac{\Delta \Omega}{4\pi}
\label{M_correct}
\eea
for large bubbles with $HR \gg 1$.

\subsection{The black-hole formation criterion}
With the corrected mass estimate~\eqref{M_correct}, 
the naive black-hole formation criterion~(\ref{eq:criterion_naive}) is now modified as
\bea
1 \lsim
\kappa H R \sqrt{\frac{\Delta\Omega}{4\pi}} 
\,.\label{eq:criterion_correct}
\eea
This is one of the main results of this work.
We did not modify our guesstimate for the circumference $C_{\cal D}$,
because $C_{\cal D}$ is measured \emph{outside} the dense region that may form a PBH
so we can still use the flat-space guesstimate of $C_{\cal D}$ as a good approximation.
Similarly, we did not modify the definition of $\Delta\Omega$ as the solid angle seen from the center in the interior of the bubble as the spacetime is approximately flat there.
Since we expect $\kappa \sqrt{\Delta\Omega/4\pi} \ll 1$, the criterion~(\ref{eq:criterion_correct}) requires $HR \gg 1$, justifying the large $HR$ approximation made above.

\subsection{Prediction of a monochromatic PBH mass spectrum}
Combining the PBH mass~(\ref{M_correct}) with the inequality~(\ref{eq:criterion_correct}), 
we get $M_\PBH \gsim \frac{8\pi}{3\kappa} H^{-3} \Delta V$.
We expect this inequality to be saturated because larger bubbles have to be born earlier but the bubble nucleation rates at earlier times are more exponentially suppressed than those for the bubbles that are barely large enough to satisfy~\eqref{eq:criterion_correct}.
Therefore, we predict that the PBH mass spectrum should be monochromatic at the value given by the threshold of the above inequality:
\bea
M_\PBH \approx \frac{8\pi}{3\kappa} H^{-3} \Delta V
\,.\label{eq:MPBH3_prediction}
\eea
%

%%%%%%%%%%%%%%%%%%%%%%%%%%%%%%%%%%%%%%%%%%%%%%%%%%%%%%%%%%%%%%%%%%%%%%%%
%%%%%%%%%%%%%%%%%%%%%%%%%%%%%%%%%%%%%%%%%%%%%%%%%%%%%%%%%%%%%%%%%%%%%%%%
\section{The PBH abundance}
\label{sec:PBHabundance}
In this section, we estimate the number density of PBHs based on our criterion~\eqref{eq:criterion_correct}.
We will show that the PBH number density $n_\PBH$ is given by 
\bea
n_\PBH 
\sim 
\int_{t_\C}^{t_{\rm f}} \hspace{-1ex} \dd t \;\e^{3H(t-t_{\rm f})}\, \Gamma_{\rm n}(t) \, P_\PBH(t)
\,,\label{result:nPBH}
\eea
where $P_\PBH(t)$ is the probability for a bubble born at time $t$ to eventually collide with another bubble and form a PBH, $\Gamma_{\rm n}(t)$ is the bubble nucleation rate per unit volume at time $t$, and $t_{\rm f}$ is the end of the phase transition. 
We will discuss the definition of $t_{\rm f}$ and the form of $P_\PBH(t)$ below, 
which will depend on the criterion~(\ref{eq:criterion_correct}) in an essential way.

\subsection{$n_\PBH$}
\label{sec:PBHnumberdensity}
Consider a bubble that was born at time $t$, and
let $P_\PBH(t)$ be the probability for this bubble to eventually form a PBH by colliding with another bubble that was born later (and hence younger and smaller). 
To express the probability for the first, older and bigger, bubble to be born,
we introduce $\Gamma_{\rm n}(t)$ as the bubble nucleation rate at time $t$ per unit volume of the false vacuum,
and $P_{\rm f}(t)$ as the co-moving 3-volume of false-vacuum regions per unit co-moving 3-volume at time $t$.
Then, with our approximation that bubbles are born at rest in the coordinates~(\ref{eq:deSitter}), the final number of PBHs within a large 3-volume ${\rm CoV}_{\!3}$ co-moving with those coordinates is given by
\bea
N_\PBH = \int_{t_\C}^{t_{\rm f}} \hspace{-1ex} \dd t \int_{{\rm CoV}_{\!3}} \hspace{-3ex} \dd^3x \,\sqrt{-g}\; P_{\rm f}(t) \, \Gamma_{\rm n}(t) \, P_\PBH(t)
\,,
\eea
where $t_{\rm f}$ marks the end of the PBH formation period (to be defined shortly), 
while $t_\C$ denotes the time corresponding to the critical temperature $T_\C$ 
marking the beginning of the phase transition.
Taking ${\rm CoV}_{\!3}$ to be sufficiently large, it is justified to disregard the ``fringe effects'' of how to count, or not count, the PBHs from the bubbles born near the boundary of ${\rm CoV}_{\!3}$.

Let us now discuss how we should like to define $t_{\rm f}$.
Since PBHs are extremely rare, we neglect the curvatures induced by the PBHs and just use the background metric for the $\sqrt{-g}$ factor.
As discussed earlier, we have a long super-cooled phase so the background metric remains de Sitter for a long time to allow for the rare and large bubbles to grow, until when a large number of small bubbles begin to form rapidly to bring the whole universe into the true vacuum.
The time $t_{\rm f}$ represents this time when small bubbles begin to fill the universe.
Clearly there is no way to define such time sharply in a unique manner, but one reasonable definition may be motivated by observing that the ratio of the volume of a sphere of diameter $\ell$ to that of a cube of side $\ell$ is given by $\pi/6 \simeq 1/2$. On average, when the volume fraction of small bubbles begin to exceed this ratio, some of the small bubbles must collide and overlap with one another, initiating the end of the phase transition (which must complete rapidly to avoid eternal inflation). We thus define $t_{\rm f}$ as
\bea
P_{\rm f}(t_{\rm f}) \equiv \frac12
\,.
\eea
With this and the approximation of taking the background metric to be de Sitter before $t_{\rm f}$,
we estimate $N_\PBH$ as
\bea
N_\PBH \sim \int_{t_\C}^{t_{\rm f}} \hspace{-1ex} \dd t \int_{{\rm CoV}_{\!3}} \hspace{-3ex} \dd^3x \;\e^{3Ht}\, \Gamma_{\rm n}(t) \, P_\PBH(t)
\, ,
\eea
where the ``$\sim$'' here represents the uncertainties coming from the subtlety in defining $t_{\rm f}$ discussed above as well as that in the precise form of the metric (the $e^{3Ht}$ factor) near $t_{\rm f}$. 
Taking the uncertainty in $t_{\rm f}$ to be $O(H^{-1})$, the total uncertainty represented by the ``$\sim$'' is $O(\e^{\pm\text{a few}})$.
To get the PBH number density $n_\PBH$, 
we divide this by the volume of ${\rm CoV}_{\!3}$ at $t=t_{\rm f}$:
\begin{equation}
\begin{aligned}
n_\PBH 
\sim\;\>&
\dfrac{\displaystyle{\int_{t_\C}^{t_{\rm f}}} \!\dd t \,\e^{3Ht}\, \Gamma_{\rm n}(t) \, P_\PBH(t)}
      {\e^{3H t_{\rm f}}} \\
=\;\>& 
\int_{t_\C}^{t_{\rm f}} \!\!\dd t \,\e^{3H(t-t_{\rm f})}\, \Gamma_{\rm n}(t) \, P_\PBH(t)
\,.
\end{aligned}
\label{eq:nPBH}
\end{equation}

In~(\ref{eq:nPBH}), $\Gamma_{\rm n}(t)$ is a quantity that should be computed in quantum field theory once the scalar sector responsible for the phase transition is specified.
Here, our task is to find $P_\PBH(t)$, namely, 
the probability for a bubble born at time $t$ to collide with another bubble born later and form a PBH\@.
Since $\Gamma_{\rm n}(t)$ is the conditional probability for a bubble to form per unit volume per unit time given that the volume is in the false vacuum 
and $P_{\rm f}(t)$ is the co-moving volume of false vacuum regions per unit co-moving volume at time $t$,
we can express $P_\PBH(t)$ in the coordinate system~(\ref{eq:deSitter}) as
\bea
\hspace{-3ex}
P_\PBH(t_1) 
= \int_{t_1}^{t_{\rm f}} \!\!\! \dd t_2 \!
  \int_{r_{\rm min}}^{r_{\rm max}} \hspace{-3ex} \dd r \; 4\pi r^2 \, \e^{3Ht_2} \, P_{\rm f}(t_{\rm coll}) \, \Gamma_{\rm n}(t_2)
\,,\label{eq:P_PBH}
\eea
where $t_1$ ($t_2$) is the time when the older (younger) bubble was born, $r$ denotes the co-moving distance between the birth location of the younger bubble and that of the older, and $t_{\rm coll}$ is the time when the two bubbles collide.
The limits $r_{\rm min}$ and $r_{\rm max}$ provide the range of $r$ in which the criterion~(\ref{eq:criterion_correct}) is satisfied, which we will derive shortly.
The presence of $P_{\rm f}(t_{\rm coll})$ discounts the PBH production if the universe is (partially) in the true vacuum at the time of the collision.
This is necessary because the bubbles grow and acquire mass by eating up the energy of the false vacuum.

One might think that ensuring no other bubbles in the backward lightcones of the two large bubbles should be extremely difficult as lots of small bubbles would keep popping up. 
But that is actually not true because $\Gamma_n$ is strongly exponentially suppressed before the nucleation time $t_{\rm n}$ defined by $\Gamma_n(t_{\rm n}) = H^4$, 
so almost all bubbles are created around or after $t_{\rm n}$, which can be seen from Fig.\,\ref{fig:benchmark_fixed_mustar} top shown later.
Therefore, once the unlikely event of forming two bubbles at early times occurred (which would eventually grow large and collide to form a PBH later), it is even more unlikely that additional bubbles would pop up between them.

\begin{figure}[t] 
\begin{center}
\includegraphics[width=0.35\textwidth]{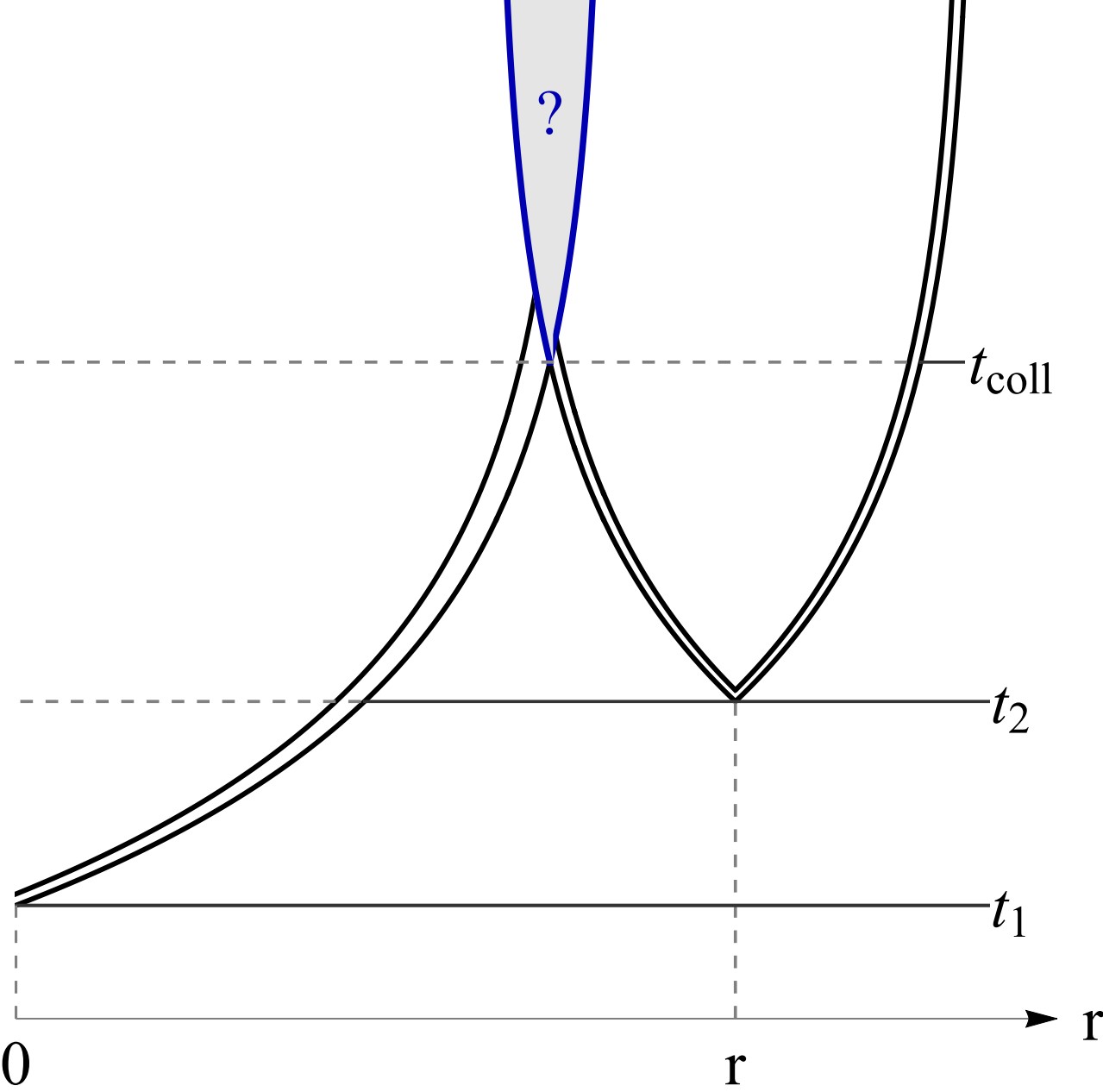} 
\end{center}
\caption{
The collision of two bubbles where with the coordinates and metric outside the bubbles are given by Eq.~(\ref{eq:deSitter}).
A PBH forms in the region with ``?'' if our criterion~(\ref{eq:criterion_correct}) is satisfied.
Constant time hypersurfaces are denoted by dashed lines inside the bubbles, as opposed to solid lines, to indicate the fact that the coordinate system used in~(\ref{eq:deSitter}) is defined and used only outside the bubbles.  
}
\label{phase_structure}
\end{figure}

Let's now find $r_{\rm min}$ and $r_{\rm max}$.
Since we are approximating bubble trajectories as lightlike, they satisfy $\dd r = e^{-Ht} \dd t$ under the metric~(\ref{eq:deSitter}). 
Then, the co-moving radii $r_1$ and $r_2$ of the older and younger bubbles at the moment of the collision are given by
\begin{equation}
\begin{aligned}
H r_1 &= \e^{-Ht_1} - \e^{-Ht_{\rm coll}}
\,,\\
H r_2 &= \e^{-Ht_2} - \e^{-Ht_{\rm coll}}
\,.
\end{aligned}
\label{eq:r1_and_r2}
\end{equation}
Then, since $r_1 + r_2 = r$ by definition,%
\footnote{These radii are not the (slant) heights of the cones in Fig.~\ref{two_bubbles_cartoon}, 
but we disregard the differences as they are $O(\Delta\Omega) \ll 1$. 
}
we find
\begin{align}
\e^{-Ht_{\rm coll}}
= \frac12 (\e^{-Ht_1} + \e^{-Ht_2} - Hr)
\,.\label{eq:collision_point}
\end{align}
Recall that the physical radius $R$ appearing in the criterion~(\ref{eq:criterion_correct}) is defined such that the surface area of a sphere of radius $R$ is given by $4\pi R^2$.
Thus, under the metric~(\ref{eq:deSitter}) outside the bubbles,
the physical radii $R_i$ corresponding to $r_i$ ($i=1,2$) at the moment of the collision are given by $R_i = r_i \,\e^{H t_{\rm coll}}$. 
Then, from~(\ref{eq:r1_and_r2}) and~(\ref{eq:collision_point}), we find
\begin{equation}
\begin{aligned}
H\! R_1 &=
\frac{\e^{-Ht_1} - \e^{-Ht_2} + Hr}{\e^{-Ht_1} + \e^{-Ht_2} - Hr} 
\,,\\
H\! R_2 &=
\frac{-\e^{-Ht_1} + \e^{-Ht_2} + Hr}{\e^{-Ht_1} + \e^{-Ht_2} - Hr} 
\,.
\end{aligned}
\label{eq:R1_and_R2}
\end{equation}
In the criterion~(\ref{eq:criterion_correct}), we assumed that the two bubbles have an equal radius $R$ for simplicity. Here, since the bubbles have different radii, we should use the smaller radius $R_2$ instead of $R$, 
and redefine $\Delta\Omega$ as the solid angle of ${\cal D}$ as seen from the center of the smaller bubble.
Therefore, with~(\ref{eq:R1_and_R2}), the criterion~(\ref{eq:criterion_correct}) can be expressed as
\begin{equation}
\begin{aligned}
&1 \lsim \kappa H\! R_2 \sqrt{\frac{\Delta \Omega}{4\pi}} 
\\
\Longleftrightarrow\quad&
(1+\kappa') \e^{-Ht_1}  +  (1-\kappa') \e^{-Ht_2} \lsim (1+\kappa') Hr
\,, 
\end{aligned}
\end{equation}
where $\kappa' \equiv \kappa\sqrt{\Delta\Omega/4\pi}$ for brevity.
Since we expect that $\kappa' \ll 1$, this is equivalent at $O(\kappa')$ to
\bea
\e^{-Ht_1}  +  (1-2\kappa') \e^{-Ht_2} \lsim Hr
\,, 
\label{eq:criterion_in_r}
\eea
which gives $r_{\rm min}$ as
\beq
Hr_{\rm min} \simeq \e^{-Ht_1} + \biggl( 1 - 2\kappa\sqrt{\frac{\Delta\Omega}{4\pi}} \biggr) \e^{-Ht_2} 
\,.\label{eq:r_min}
\eeq
On the other hand, $r_{\rm max}$ can be found from~(\ref{eq:collision_point}) by taking $t_{\rm coll}$ in~(\ref{eq:collision_point}) to infinity:
\beq
Hr_{\rm max} \simeq \e^{-Ht_1} + \e^{-Ht_2} 
\,.\label{eq:r_max}
\eeq
One might object to the $t_{\rm coll} \to \infty$ limit because the de Sitter expansion does not last forever as the phase transition ends and the universe becomes radiation dominated.
In our treatment this ``cutoff'' is effectively provided by $P_{\rm f}(t_{\rm coll})$ in~(\ref{eq:P_PBH}).

\subsection{$P_{\rm f}(t)$}
The last remaining ingredient we need to complete the expression~(\ref{eq:P_PBH}) is $P_{\rm f}(t)$,
the co-moving volume of false-vacuum regions per unit co-moving volume at time $t$.
We can obtain $P_{\rm f}(\tau)$, where $\tau$ is the conformal time, as follows.
Between the times $\tau'$ and $\tau' + \dd\tau'$, the number of newly nucleated bubbles per unit co-moving volume is given by $\Gamma_{\rm n}(\tau') \cdot a(\tau')^4 \, P_{\rm f}(\tau') \, \dd\tau'$.
After the time $\tau'$, in any interval between $\tau''$ and $\tau'' + \dd\tau''$, the bubbles born at $\tau'$ grow by ``eating'' part of the false vacuum. 
The eaten co-moving volume is given by $ P_{\rm f}(\tau'') \cdot 4\pi r(\tau'', \tau')^2\, \dd r$, where $r(\tau'', \tau')$ is the co-moving radius at time $\tau''$ of a bubble born at time $\tau'$, $\dd r$ is the growth in $r(\tau'', \tau')$ during the interval $\dd\tau''$,
and the factor of $P_{\rm f}(\tau'')$ is there because the bubbles cannot eat a volume that is already in the true vacuum.
Therefore, we obtain
\begin{equation}
\begin{aligned}
P_{\rm f}(\tau) 
= 
1-
&\!\int_{\tau_\C}^{\tau} \!\! d\tau' \,
P_{\rm f}(\tau') \, a(\tau')^4 \,  \Gamma_{\rm n}(\tau') \>\times\\
&\times\!\int_{\tau'}^{\tau} \!\! d\tau'' \,
P_{\rm f}(\tau'') \cdot 
4\pi r(\tau'', \tau')^2\, \frac{\partial r(\tau'',\tau')}{\partial\tau''}
\,,
\end{aligned}
\label{eq:P_f}
\end{equation}
where, since we approximate the bubble wall trajectories to be lightlike, we have
\bea
r(\tau'',\tau') \simeq \tau''-\tau'
\quad\text{and}\quad
\frac{\partial r(\tau'',\tau')}{\partial\tau''} \simeq 1
\,.
\eea
Furthermore, we also approximate that $a(\tau)$ is given by the de Sitter expansion, i.e.,
\bea
a(\tau) = \frac{a_0}{1 - Ha_0(\tau - \tau_0)}   
\,,
\eea
where for convenience we choose the convention that $\tau_0 = \tau_{\rm n}$ (the nucleation time) and $a_0 = 1$, although the results are convention independent.

%%%%%%%%%%%%%%%%%%%%%%%%%%%%%%%%%%%%%%%%%%%%%%%%%%%%%%%%%%%%%%%%
%%%%%%%%%%%%%%%%%%%%%%%%%%%%%%%%%%%%%%%%%%%%%%%%%%%%%%%%%%%%%%%%
\subsection{$M_\PBH$ and $f_\PBH$}
\label{sec:M_PBH-and-f_PBH}
We would first like to rewrite the prediction of $M_\PBH$ in Eq.~\eqref{eq:MPBH3_prediction} 
by eliminating $H$ from it using $H^2 = 8\pi \Delta V / (3\Mpl^2)$ with $\Mpl = 1.22\times 10^{19}\>\GeV$, and then eliminating $\Delta V$ in favor of the highest possible reheating temperature $\Tstarmax$ defined via $\Delta V = \frac{\pi^2}{30} g_{*{\rm SM}} \Tstarmax^4$. This leads to
\bea
M_\PBH 
= \frac{\Mpl^3}{\kappa} \sqrt{\frac{3}{8\pi\Delta V}}
= \frac{1}{\kappa} \sqrt{\frac{45}{4\pi^3 g_{*{\rm SM}}}} \frac{\Mpl^3}{\Tstarmax^2}
\,.\label{eq:MPBHinTstar}
\eea
For the PBH number density, it is the ratio $n_\PBH / H^3$ that is most easily obtainable from the method described in Sec.~\ref{sec:PBHnumberdensity} by working in the units where $H=1$. 
Then, multiplying the right-hand side of Eq.~\eqref{eq:MPBH3_prediction} by $n_\PBH$ and combining the $H^{-3}$ with $n_\PBH$, and then eliminating the $\Delta V$ in favor of $\Tstarmax$, we get
\begin{equation}
\begin{aligned}
f_\PBH 
&\equiv \frac{M_\PBH \, n_\PBH}{\rho_\DM} \\
&= \frac{8\pi}{3\kappa} \frac{\frac{\pi^2}{30} g_{*{\rm SM}} \Tstarmax^4}{\rho_\DM} \, \frac{n_\PBH}{H^3} \\
&= \frac{2\pi}{\kappa} \frac{\Tstarmax}{\rho_\DM/s_*} \!\left( \frac{\Tstarmax}{T_*} \right)^{\!\!3} \frac{g_{*{\rm SM}}}{g_*} \frac{n_\PBH}{H^3}
\,,
\end{aligned}
\label{eq:fPBHinTstar}
\end{equation}
where we have introduced the entropy density $s_* = (2\pi^2/45) g_* T_*^3$ at the reheating, 
and the value of $\rho_\DM/s_*$ is $\simeq 0.44\>{\rm eV}$ from observation.

\subsection{The nucleation rate and a benchmark microscopic model}
Now we have all ingredients necessary to obtain $n_\PBH$ once the nucleation rate $\Gamma_{\rm n}(t)$ is provided.
For the reader familiar with the literature, we should mention that we \emph{cannot} adopt the commonly used parametrization for $\Gamma_{\rm n}$ proposed by~\cite{Turner:1992tz, Kamionkowski:1993fg}: $\Gamma_{\rm n}(t) = H^4 \e^{\beta (t-t_{\rm n})}$,
where $\beta$ is a free parameter and $t_{\rm n}$ the nucleation time.
The reason is the following.
This parametrization is essentially a Taylor expansion around $t=t_{\rm n}$ in the exponent, 
so it is a good model-independent parametrization if $t$ is sufficiently near $t_{\rm n}$.
But we are interested in large bubbles that were born much earlier than $t_{\rm n}$, 
so the Taylor expansion is not justified.
Since there is no good model-independent way to parametrize $\Gamma_{\rm n}(t)$ at such early times, 
we must choose a benchmark model.
We adopt a simple Abelian Higgs model as our benchmark because, as we already mentioned, 
a thick fluid shell may be realized by a phase transition that breaks gauge symmetry, 
and the phase transition can be made strongly first order.
While the details of the benchmark model is in Appendix~\ref{app:lagrangian}, 
what is essential for the discussions below is that the model has two free parameters, the $U(1)$ gauge coupling $g$ and a mass parameter $\mu_*$, which is the gauge boson mass up to an $O(1)$ numerical factor.

The bubble nucleation rate can be written as
\bea
\Gamma_{\rm n}(T) = T^4 \, \e^{-E[\phi_\C]/T}
\, ,
\eea
where $E[\phi_\C]$ is the energy of the so-called ``critical bubble'' configuration and we have suppressed the numerical factor in front of $T^4$ (see Appendix~\ref{app:nucleation_rate} for details).    
We define the nucleation temperature $T_{\rm n}$ via $\Gamma_{\rm n}(T_{\rm n}) = H^4$, 
where $H^2 = 8\pi\Delta V / (3M_{\rm Pl}^2)$, 
neglecting the energy density in radiation compared to $\Delta V$, 
which is a good approximation in a supercooled phase.
This definition can then be inverted to express the required value of $E[\phi_\C]/T$ for the bubble nucleation rate to match the expansion rate at $T=T_{\rm n}$:
\begin{equation}
\frac{E[\phi_\C]}{T} \biggr|_{T = T_{\rm n}} \!\!\!\!\!\!\!
=\> \ln \!\left[ \!\left( \frac{3M_{\rm Pl}^2 T_{\rm n}^2}{8\pi\Delta V} \right)^{\!\!2} \right]\! 
\>\simeq\>
6.5 + 4\ln\!\left[ \frac{gT_{\rm n}}{\mu_*} \frac{gM_{\rm Pl}}{\mu_*} \right]\!
\,,\label{Tn_condition}
\end{equation}
where $M_{\rm Pl} = 1.22 \times 10^{19}\,\GeV$.

\begin{figure}[t] 
\begin{center}
\includegraphics[width=0.45\textwidth]{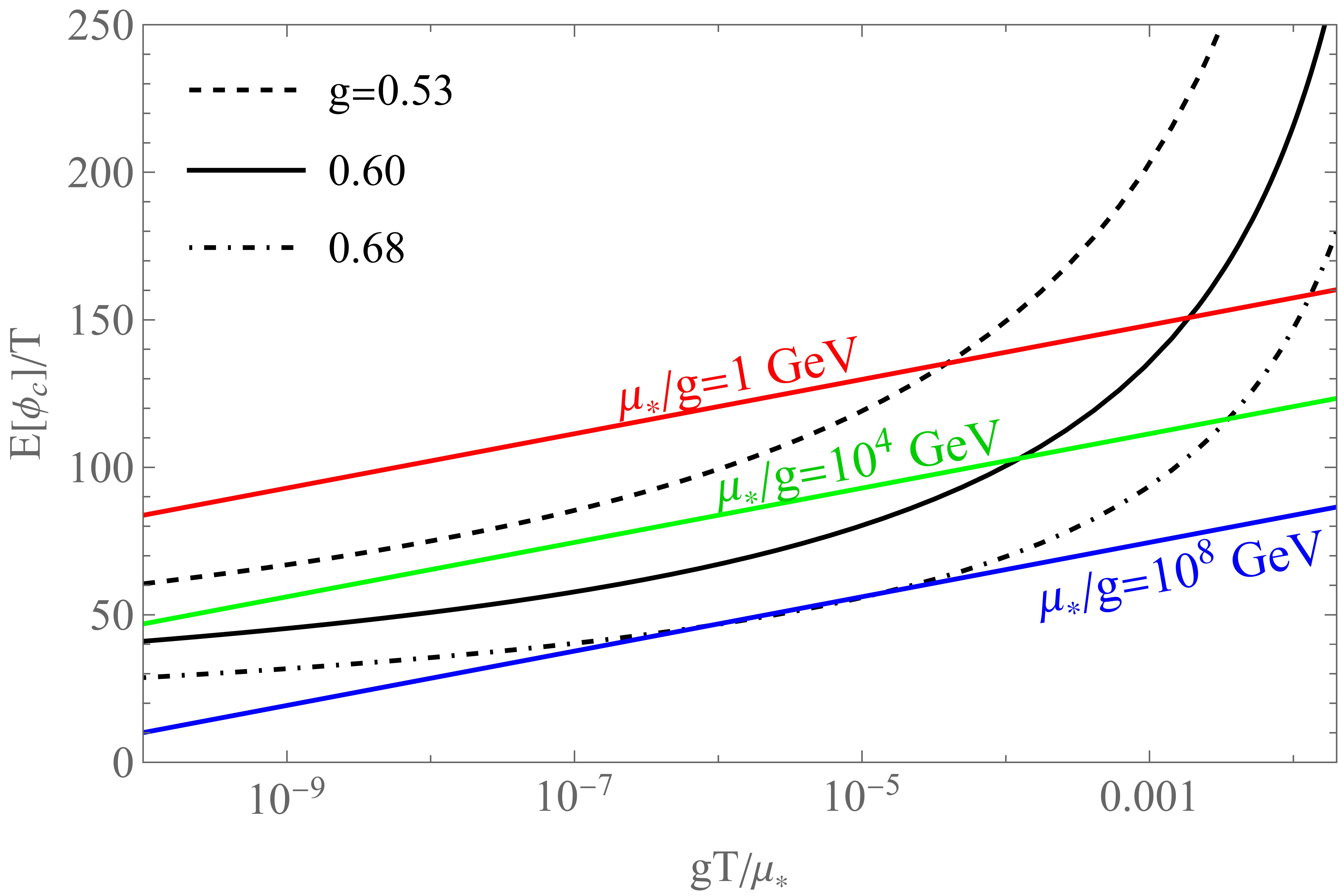} 
\end{center}
\caption{The black curves show $E[\phi_\C]/T$ for $g=0.53$ (dashed), $0.60$ (solid), and $0.68$ (dot-dashed) as a function of $gT/\mu_*$. The straight lines in different colors show the required values of $E[\phi_\C]/T$ for the bubble nucleation rate to match the expansion rate of the universe when $\mu_*/g = 1\,\GeV$ (red), $10^4\,\GeV$ (green), and $10^8\,\GeV$ (blue). For given value of $\mu_*/g$, the nucleation rate is larger (smaller) than the expansion rate when the curve is below (above) the line.}
\label{fig:S3overT}
\end{figure}

In Fig.~\ref{fig:S3overT}, the black curves show $E[\phi_\C]/T$ as a function of $gT / \mu_*$ for different values of $g$,
while the straight lines show the value of the right-hand side of~(\ref{Tn_condition}) as a function of $gT/\mu_*$ for different values of $\mu_*/g$.
For each value of $g$, 
as we lower the temperature and follow its black curve from top-right to lower-left, 
when the curve meets the line of given $\mu_*/g$, that intersection gives $T_{\rm n}$ for that $\mu_*/g$.
Since the right-hand side of~(\ref{Tn_condition}) says that the vertical offset of the line increases as $\mu_*/g$ decreases, the curve is guaranteed to intersect with the line for any given $g$ if we choose $\mu_*/g$ to be sufficiently low.
When $\mu_*/g$ gets too high, no intersections exist and the bubble nucleation rate would never be large enough to catch up with the expansion rate of the universe.
This feature was also pointed out in Ref.\,\cite{Ellis:2018mja, Lewicki:2021xku}.

To use $\Gamma_{\rm n}$ for the estimation of the PBH abundance described in the previous sections, 
we need $\Gamma_{\rm n}$ as a function of $t$ rather than $T$. 
To relate them we use $T(t) = T_{\rm n} \, \e^{-H(t-t_{\rm n})}$. 
This should be a good approximation as long as the universe is dominated by the vacuum energy, which is the case during the bubble formation period except near the very end of the phase transition when the spacetime is no longer approximately de Sitter.

\begin{figure}[h] 
\begin{center}
\includegraphics[width=0.46\textwidth]{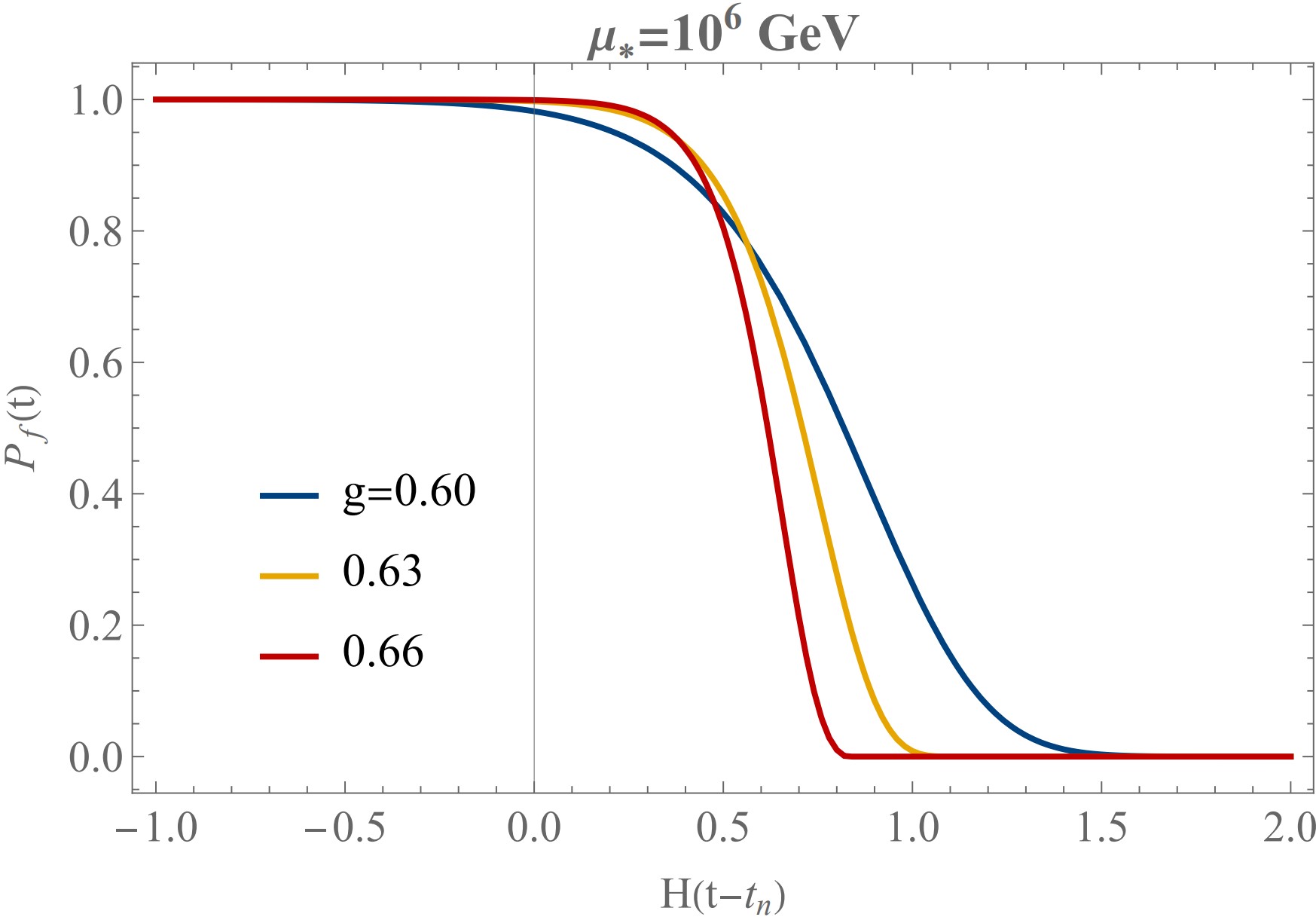} 
\includegraphics[width=0.46\textwidth]{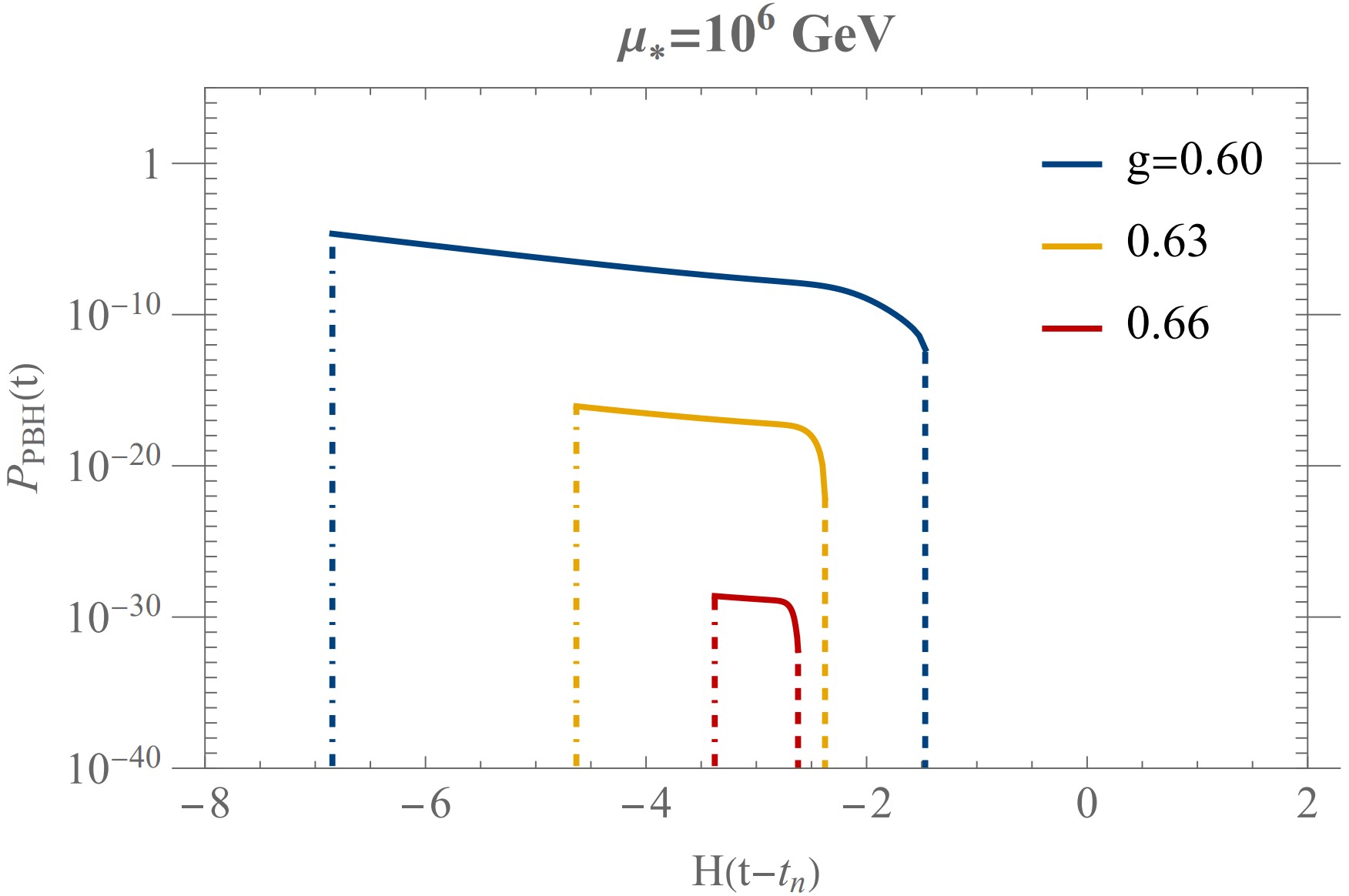} 
\includegraphics[width=0.46\textwidth]{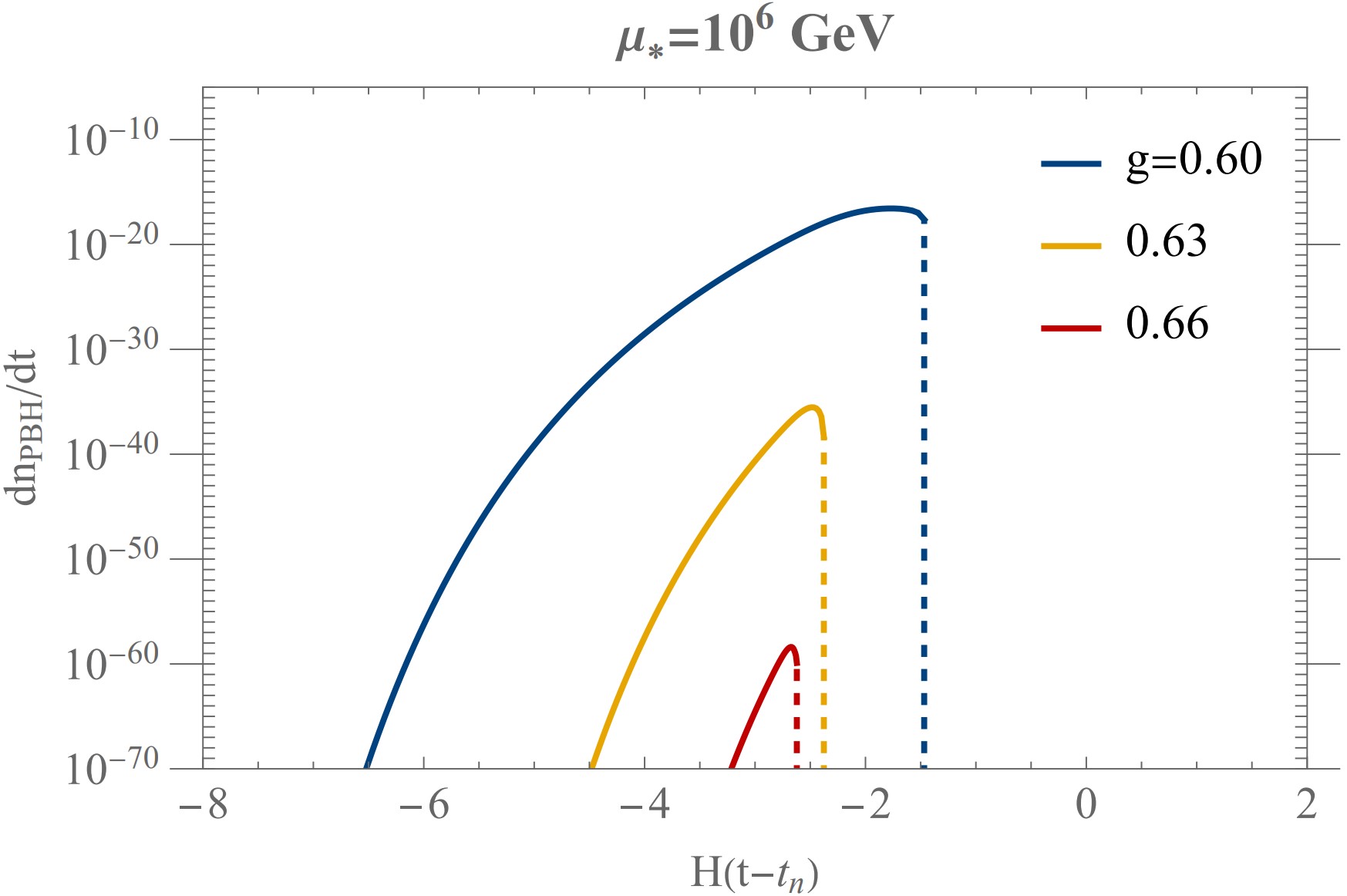} 
\end{center}
\caption{$P_{\rm f}(t)$ (top), $P_\PBH(t)$ (middle), and $\dd n_\PBH / \dd t$ (bottom) 
as a function of $H(t-t_{\rm n})$ with $\kappa=1$ and $\Delta\Omega/4\pi=10^{-3}$, 
where $\dd n_\PBH / \dd t$ is the integrand of the right-most expression in~(\ref{eq:nPBH}).
The dashed lines in the middle and bottom plots indicate the fact that $P_f(t_{\rm coll})$ effectively goes to zero in our numerical accuracy.
The dot-dashed lines in the middle plot correspond to $t_{\rm c}$.}
\label{fig:benchmark_fixed_mustar}
\end{figure}

Once $\Gamma_{\rm n}(t)$ is thus obtained, we calculate $P_{\rm f}(t)$, $P_\PBH(t)$, and $n_\PBH$ by~\eqref{eq:P_f}, \eqref{eq:P_PBH}, and \eqref{eq:nPBH}, respectively. 
In Fig.~\ref{fig:benchmark_fixed_mustar}, $P_{\rm f}(t)$, $P_\PBH(t)$, and the \emph{integrand} of $n_\PBH$ are plotted for the benchmark values of $\mu_* = 10^6\,\GeV$ and three different values of $g$, all with $\kappa=1$ and $\Delta\Omega/4\pi=10^{-3}$.

\begin{figure*}[t] 
\begin{center}
\includegraphics[width=0.48\textwidth]{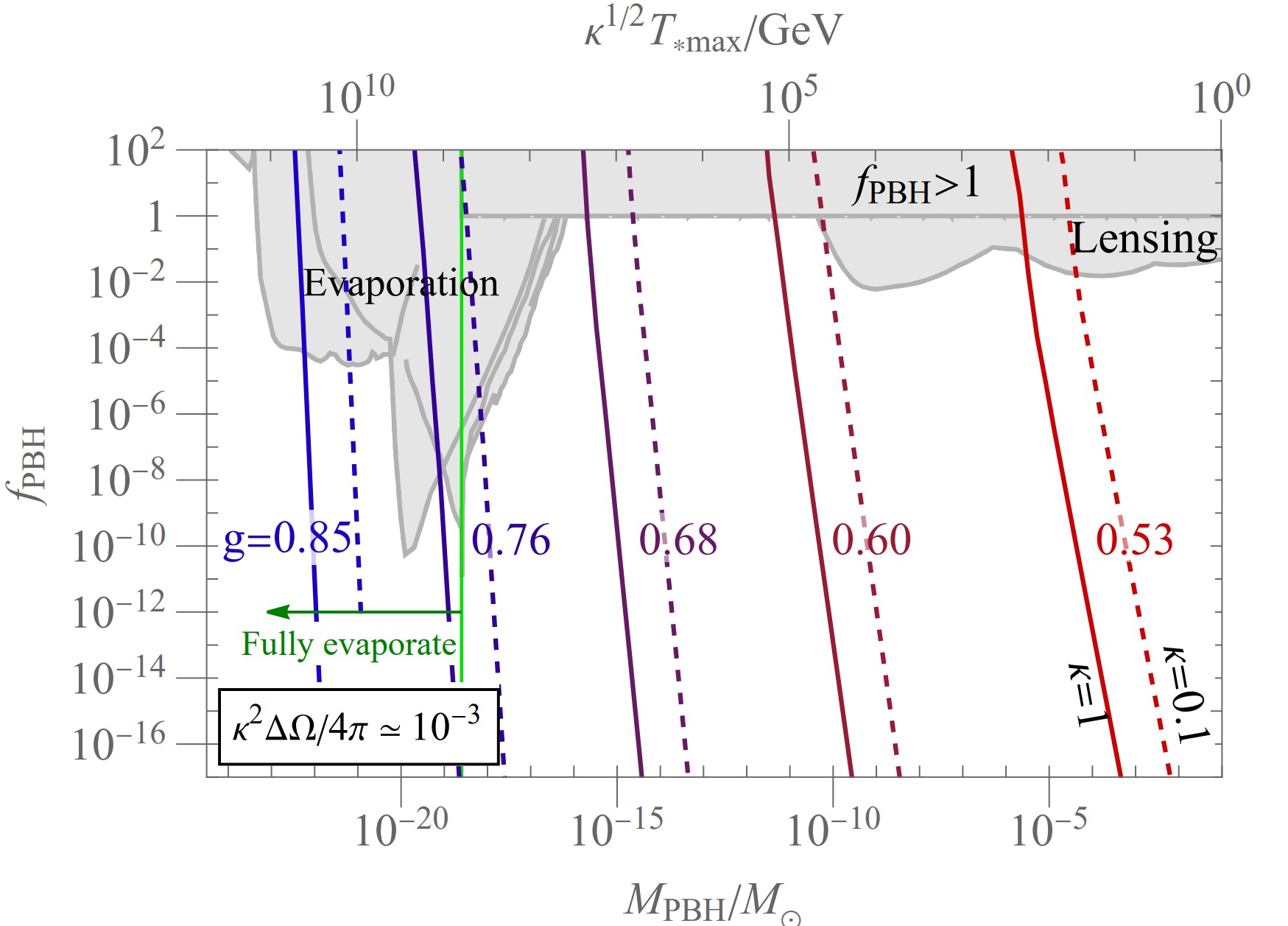} 
\includegraphics[width=0.48\textwidth]{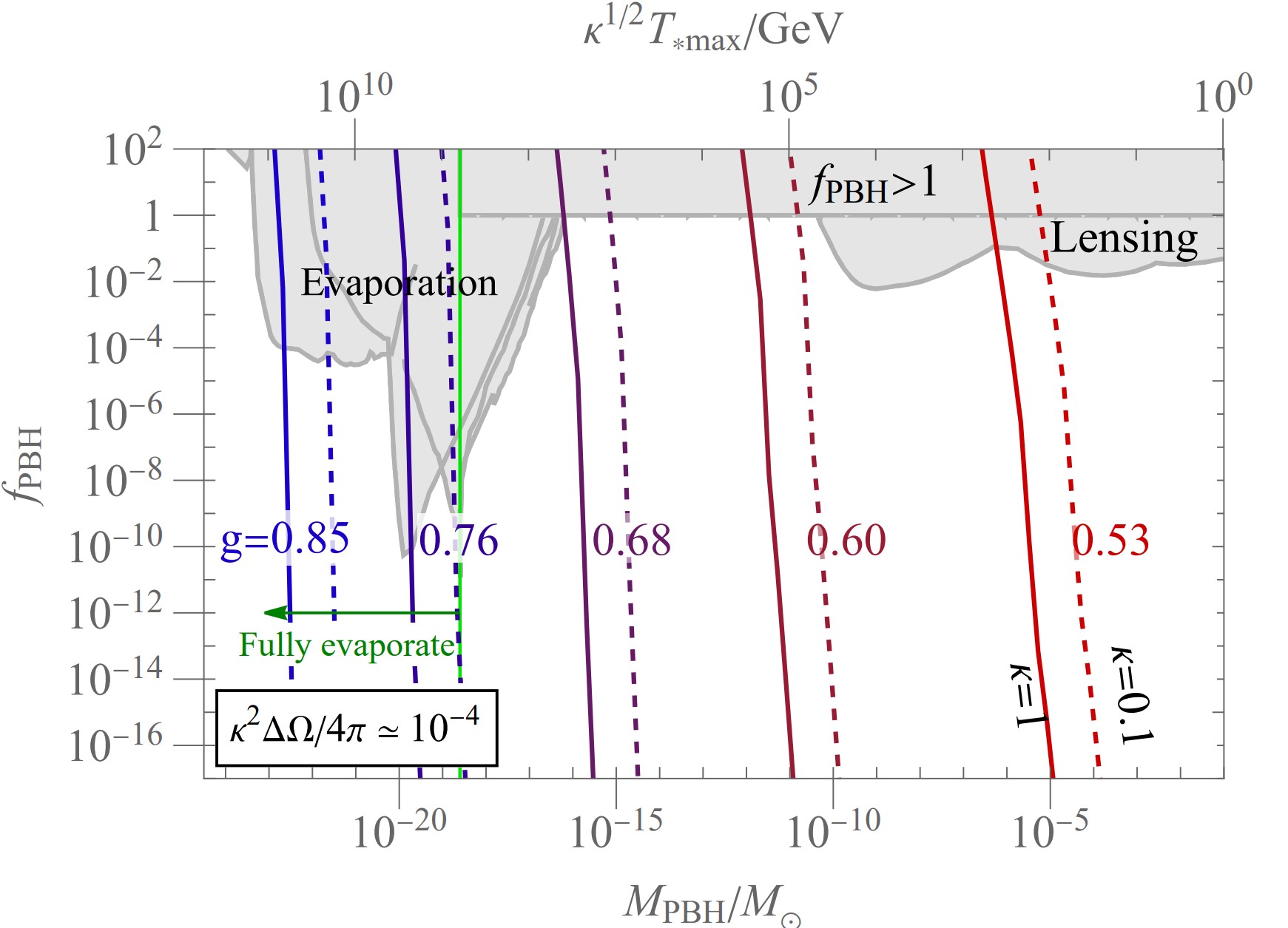} 
\end{center}
\caption{The predicted relation between the PBH abundance $f_\PBH$ and $M_\PBH$ (bottom axis) or $\sqrt{\kappa} \,\Tstarmax$ (top axis) 
for $\kappa^2\Delta \Omega/4\pi = 10^{-3}$ (left), and $10^{-4}$ (right).
The relation further depends on the U(1) gauge coupling $g$ and the parameter $\kappa$.
Different values of $g$ are represented by lines of different colors, where for each color two different values of $\kappa$ are represented by solid ($\kappa=1$) and dashed ($\kappa=0.1$) lines.
These results are for $T_* = \Tstarmax$ and $g_* = g_{*{\rm SM}}$. 
The $T_*$ and $g_*$ dependencies can be obtained by simple rescaling.
Current observational constraints, taken from~\cite{Carr:2020gox, Carr:2020xqk}, are depicted by gray shaded regions.
The vertical axis in each pane denotes the values of $f_\PBH$ that we would observe today if we ignored Hawking evaporation, and the PBHs lighter than $M_\PBH = 5 \times 10^{14}\>{\rm g}$~\cite{Carr:2009jm, Carr:2016hva} (depicted by the green solid vertical line) would actually have fully evaporated by today.}
\label{fig:result}
\end{figure*}
%
 
%%%%%%%%%%%%%%%%%%%%%%%%%%%%%%%%%%%%%%%%%%%%%%%%%%%%%%%%%%%%%%%%%%%%%%%%
%%%%%%%%%%%%%%%%%%%%%%%%%%%%%%%%%%%%%%%%%%%%%%%%%%%%%%%%%%%%%%%%%%%%%%%%

\section{Results}
Next, we treat $\kappa$ and $\kappa^2\Delta\Omega$ as free parameters because we would need detailed numerical relativity+fluid dynamics simulations to calculate them, which we save for future work. 
The reason for treating the combination $\kappa^2\Delta\Omega$ as an independent parameter rather than $\Delta\Omega$ itself is because the PBH formation criterion~(\ref{eq:criterion_correct}) depends only on the combination $\kappa^2\Delta\Omega$ and not on $\kappa$ or $\Delta\Omega$ separately, so the PBH \emph{number} density depends only on $\kappa^2\Delta\Omega$.
On the other hand, the PBH mass depends only on $\kappa$ in~(\ref{eq:MPBH3_prediction}).
There is still a potentially serious complication that in reality $\kappa$ and $\kappa^2\Delta\Omega$ are likely to depend also on the quantities that vary collision-by-collision like the time elapsed since $t_{\rm n}$, the bubble radii at the time of the collision, etc.
In this work we ignore this complication and use the same values of $\kappa$ and $\kappa^2\Delta\Omega$ for all collisions.

Lastly, the reheating temperature $T_*$ is also a free parameter.
Reheating occurs as the phase transition ends by a large number of small bubbles colliding and filling up the universe with the true vacuum, during which the energy of the bubble shells, which were originally the false vacuum energy, are being converted to heat in the plasma permeating in the true vacuum. Since this end stage of the phase transition is not instantaneous but lasts for a time of $O(H^{-1})$, not 100\% of the original energy of the false vacuum would be converted to heat but some fraction of it would be lost to the expansion of the universe.
The value of $T_*$ also depends on $g_*$, the radiation degrees of freedom at $T = T_*$, which is necessarily larger than that of the SM, $g_{*{\rm SM}} = 106.75$.
Therefore, we have $T_* < \Tstarmax$, where $\Tstarmax$ is defined via $\frac{\pi^2}{30} g_{*\rm SM} \Tstarmax^4 = \Delta V$.
Since the value of $T_*$ depends on details of the model and reheating dynamics, in this work we treat $T_*$ as a free parameter bounded by $T_{*{\rm max}}$.

With these limitations in mind, we show in Fig.~\ref{fig:result} our results for the PBH abundance $f_\PBH$ vs the PBH mass $M_\PBH$ for several benchmark values of $\kappa^2\Delta\Omega/4\pi$.
Since $M_\PBH$ and $\sqrt{\kappa}\,\Tstarmax$ has a one-to-one correspondence (see Sec.~\ref{sec:M_PBH-and-f_PBH}), 
we also show the values of $\sqrt{\kappa}\,\Tstarmax$ in the upper horizontal axes.
At given $M_\PBH$ or $\sqrt{\kappa}\,\Tstarmax$, 
the abundance depends on the U(1) gauge coupling constant $g$ of the Abelian Higgs benchmark model and also on $\kappa$.
Different values of $g$ are distinguished by lines of different colors, while different values of $\kappa$ are shown in lines of different forms (solid or dashed).
To avoid too much clatter in the plots, we show the results only for $T_* = \Tstarmax$, and $g_* = g_{*{\rm SM}}$. 
There is no real loss of generality, however, because both $M_\PBH$ and $f_\PBH$ scale very simply in terms of $g_*/g_{*{\rm SM}}$ and $T_*/\Tstarmax$. 
As shown in Sec.~\ref{sec:M_PBH-and-f_PBH}, 
$f_\PBH$ should be multiplied by $(\Tstarmax / T_*)^3 (g_{*{\rm SM}} / g_*)$. 
As discussed in Sec.~\ref{sec:PBHnumberdensity}, 
the dependences of $f_\PBH$ on $\kappa$ and $\kappa^2\Delta\Omega$ are very complicated, 
and in particular it is not possible to express them by simple re-scalings.

The vertical axes in Fig.~\ref{fig:result} represent the values of $f_\PBH$ that we would observe if we ignored the evaporation of the PBHs by Hawking radiation~\cite{Hawking:1974rv, Hawking:1975vcx}.
As discussed in~\cite{Carr:2009jm, Carr:2016hva}, the PBHs lighter than about $5\times 10^{14}\,{\rm g}$ (indicated by the green solid vertical line in each pane of Fig.~\ref{fig:result}) would have fully evaporated by today.

The gray regions in Fig.\,\ref{fig:result} represent constraints on $f_\PBH$ taken from \cite{Carr:2020gox, Carr:2020xqk}. 
The PBH abundance in the mass range of $10^{-10}M_{\odot} \lsim M_\PBH \lsim 10 M_{\odot}$, corresponding to $0.1\,\GeV \lsim \sqrt{\kappa}\,\Tstarmax \lsim 10^5\,\GeV$, is constrained by gravitational lensing~\cite{Marani:1998sh, Nemiroff:2001bp,Barnacka:2012bm, Katz:2018zrn, Griest:2013esa, Griest:2013aaa, Niikura:2017zjd, Paczynski:1985jf, Hamadache:2006fw, Wyrzykowski:2009ep, CalchiNovati:2009kq, Wyrzykowski:2010bh, Wyrzykowski:2010mh, Wyrzykowski:2011tr, Niikura:2019kqi, Zumalacarregui:2017qqd, Garcia-Bellido:2017imq}, 
while the range below $10^{-16} M_{\odot}$, corresponding to $\sqrt{\kappa}\,\Tstarmax \gsim 10^8\,\GeV$, is strongly constrained by Hawking evaporation as those PBHs would have partially or fully evaporated with observable signals in various channels~\cite{Page:1976wx, Carr:2009jm, Carr:2016hva, Boudaud:2018hqb, DeRocco:2019fjq, Laha:2019ssq, Dasgupta:2019cae, Laha:2020ivk, Chan:2020zry, Belotsky:2014twa, Kim:2020ngi}.
Those in the range $M_\PBH \lsim 10^{-19} M_\odot$ (or $\sqrt{\kappa}\,\Tstarmax \gsim 10^9\,\GeV$) on the left of the green solid line would have completely evaporated by today~\cite{Carr:2009jm, Carr:2016hva} so they could not constitute dark matter even if we disregarded the evaporation signals.

\begin{figure}[t] 
\begin{center}
\includegraphics[width=0.48\textwidth]{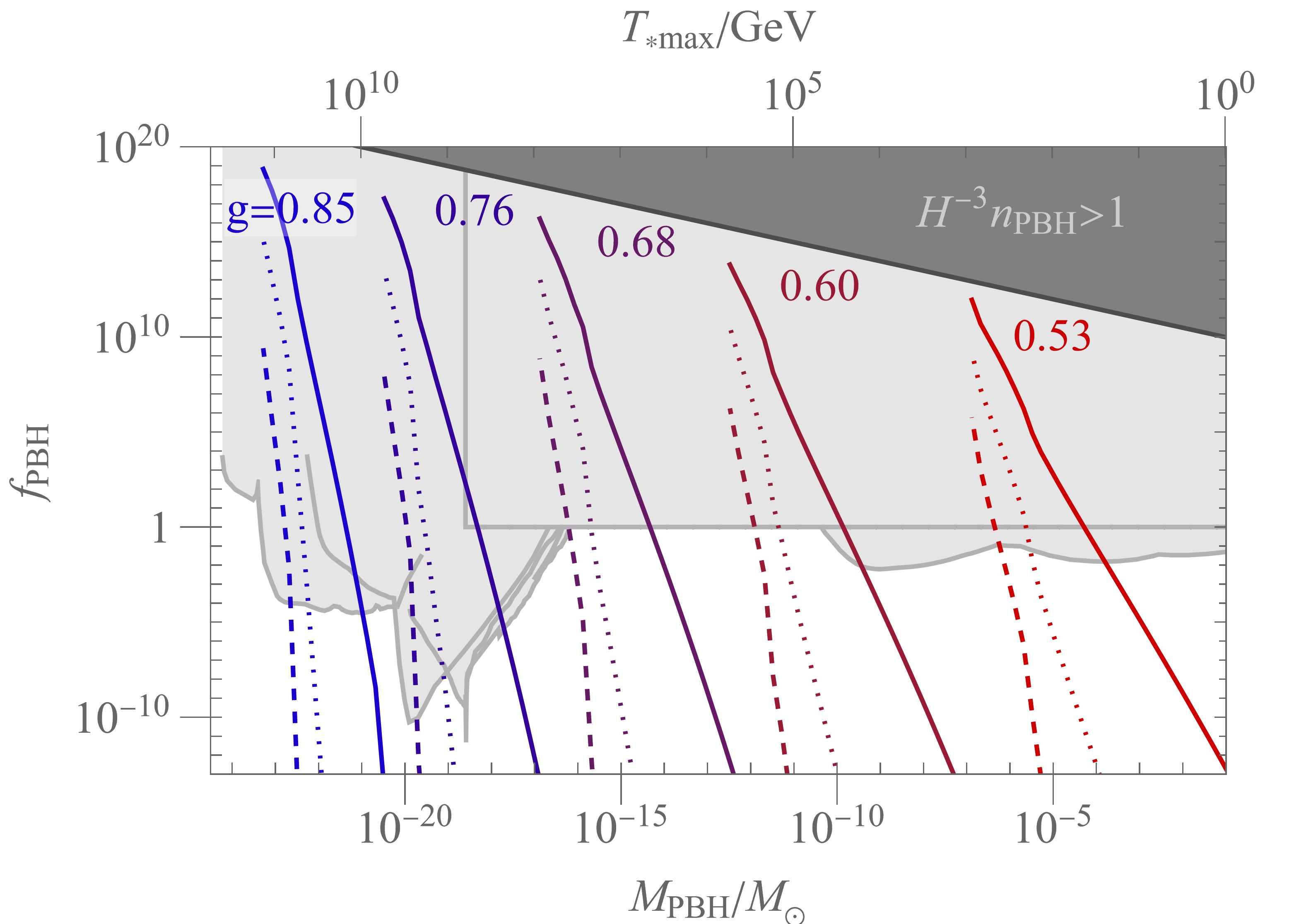} 
\end{center}
\caption{The PBH abundance $f_\PBH$ versus $M_\PBH$ (bottom axis) or $\Tstarmax$ (top axis)
for $\kappa=1$ and various values of $g$ represented by different colors.
For each color, the solid, dotted, and dashed lines correspond to $\Delta \Omega/4\pi=10^{-2}$, $10^{-3}$ and $10^{-4}$, respectively.}
\label{fig:zoomed_out}
\end{figure}

In Fig.~\ref{fig:zoomed_out}, we put our results in a larger plot range,
where different values of $g$ are represented by different colors, 
and for each value of $g$, the solid, dotted, and dashed curves correspond to $\Delta \Omega/4\pi=10^{-2}$, $10^{-3}$ and $10^{-4}$, respectively, 
all with $\kappa=1$.
In particular, we can see in this plot how the sensitivity of $f_\PBH$ on the values of $g$ and $\mu_*$ rapidly increases as $\Delta \Omega/4\pi$ is reduced, reflecting the fact that
$f_\PBH$ gets exponentially suppressed as $\Delta\Omega$ decreases.
Notice that all the curves terminate as we follow them in the upper-left direction. 
Those endpoints correspond to the values of $\mu_*/g$ beyond which the two intersections of the line with the black curve in Fig.~\ref{fig:S3overT} are too close to each other to have a long enough bubble nucleation period.

The most interesting range for us is $10^{-16} M_\odot \lsim M_\PBH \lsim 10^{-10} M_\odot$, corresponding to $10^5\,\GeV \lsim \sqrt{\kappa}\,\Tstarmax \lsim 10^8\,\GeV$, 
which is currently unconstrained by any PBH searches. 
The value of the U(1) gauge coupling $g$ only needs to be mildly tuned to 10\% level to have $f_\PBH = 1$ in this mass window.
But as we can see from the steepness of the colored lines in Fig.~\ref{fig:result}, the value of $f_\PBH$ for given $g$ is exponentially sensitive to $M_\PBH$ or the overall mass scale of the theory.
This seems inevitable in our scenario not only because our PBH production is due to large bubbles that were born long before $t_{\rm n}$ and hence extremely rare, but also because the rarity itself, $\Gamma_{\rm n}$, is exponentially sensitive to the underlying parameters of the theory.    
Therefore, we conclude that%
\begin{itemize}
\item{Dark matter may be (entirely) comprised of PBHs of a monochromatic mass spectrum produced from bubble collisions during a cosmological first-order phase transition in a rather mundane field theory like a U(1) gauge theory with an $O(1)$ gauge coupling spontaneously broken at an intermediate energy scale above the electroweak scale.}
\item{At the same time, it is highly non-generic to land near $f_\PBH = 1$ and/or in the allowed window of $M_\PBH$, so PBH production from bubble collisions can put severe constraints on field-theoretic models with cosmological first-order phase transitions.}
\end{itemize}
These are the main messages of our work.

%

%%%%%%%%%%%%%%%%%%%%%%%%%%%%%%%%%%%%%%%%%%%%%%%%%%%%%%%%%%%%%%%%%%%%%%%%
%%%%%%%%%%%%%%%%%%%%%%%%%%%%%%%%%%%%%%%%%%%%%%%%%%%%%%%%%%%%%%%%%%%%%%%%
\section{Prospect for future work}
While we have shown that bubble collisions during a first-order phase transition could produce a sizable abundance of PBHs or even over-close the universe, there are a few important loose ends that need to be worked out in future work.

(1): The first important problem is to understand the dynamics of a single, large bubble, especially the friction on the expanding bubble wall and the thickness of the fluid shell, because a macroscopically thick fluid shell is necessary for PBH production.
The challenge is to study this in a curved spacetime because our bubbles are large.

(2): Next step will be to perform a numerical relativity simulation of collision of two large bubbles to get the parameters $\kappa$ and $\kappa^2\Delta\Omega$
to be used for our parametrized criterion\,\eqref{eq:criterion_correct}.
We expect that 
while the thickness of each fluid shell is certainly an important factor, 
the most important thing is the self-interaction strength of the plasma 
so that the two shells \emph{collide} rather than passing each other to form a hot and dense region.

For those studies, numerical relativity simulations with the adopted mesh refinement (AMR), e.g., GRChombo\,\cite{Clough:2015sqa,Andrade:2021rbd}, seem to be useful as they can optimize computing resources by dynamically distributing more lattice points near bubble walls while maintaining sparse points inside or outside the bubbles where the geometry is almost flat or de-Sitter.
Since a two-bubble system possesses $SO(2,1)$ symmetry, a full (3+1)-dimensional simulation is unnecessary.
However, we expect that for realistic values of $\mu_*$ and $g$ with $\mu_* \ll M_{\rm Pl}$ and $g \lsim 1$, the scalar wall widths will likely turn out to be too small for the AMR technique to adapt, so we will need to explore how we may extrapolate the results from unrealistic values.

(3):
It will be also interesting and important to simulate collisions of many small bubbles to study reheating.
Recall that PBHs are produced from rare collisions of large bubbles, while nearly all collisions are collisions of small bubbles happening near the end of the phase transition, which reheat the universe.
The question is how close $T_*$ is to $\Tstarmax$.

(4) Another direction is to explore other benchmarks. 
While our U(1) benchmark model appears to be one of the simplest possible models,  
it is completely an independent module and not connected to the Standard Model.
Perhaps it would be more satisfactory if it is more integrated into a bigger structure including the Standard Model and other crucial elements such as inflation. 

As for PBH signals and constraints, 
the currently open region between $10^{-16} M_\odot$ and $10^{-10} M_\odot$ may be partly probed through Hawking evaporation by future {\rm MeV} gamma-ray telescopes \cite{Coogan:2020tuf, Ray:2021mxu, Ghosh:2021gfa} such as
AdEPT~\cite{Hunter:2013wla}, AMEGO~\cite{AMEGO:2019gny},
eASTROGAM~\cite{e-ASTROGAM:2017pxr},
GECCO~\cite{GECCO}, MAST~\,\cite{Dzhatdoev:2019kay},
PANGU~\cite{Wu:2014tya}, GRAMS~\cite{Aramaki:2019bpi} and XGIS~\cite{Labanti:2021gji, THESEUSConsortium:2021yvf}, 
as well as through a future extension of the analysis of Ref.~\cite{Niikura:2017zjd} with more data.
This region may also be covered by future gravitational wave experiments because first-order phase transitions with $t_{\rm f} - t_{\rm n}$ comparable to $H^{-1}$ like in our scenario can enhance gravitational wave production~\cite{Caprini:2019egz, Schmitz:2020syl}.
Moreover, such phase transition can lead to a distribution of bubble radii that may give rise to distinct features in the gravitational wave spectrum. 
We leave these interesting questions for future work.

%%%%%%%%%%%%%%%%%%%%%%%%%%%%%%%%%%%%%%%%%%%%%%%%%%%%%%%%%%%%%%%%%%%%%%%%
%%%%%%%%%%%%%%%%%%%%%%%%%%%%%%%%%%%%%%%%%%%%%%%%%%%%%%%%%%%%%%%%%%%%%%%%
\section*{acknowledgment}
THJ thanks Chang Sub Shin for helpful discussions regarding the hoop conjecture.
We also thank Kiyoharu Kawana for his valuable feedback and discussions on our earlier version of the paper.
This work is supported by the US Department of Energy (DOE) grant DE-SC0010102, 
by the Japan Society for Promotion of Science (JSPS) grant KAKENHI 21H01086,
and by the Institute for Basic Science (IBS) under the project code IBS-R018-D1.

%%%%%%%%%%%%%%%%%%%%%%%%%%%%%%%%%%%%%%%%%%%%%%%%%%%%%%%%%%%%%%%%%%%%%%%%
%%%%%%%%%%%%%%%%%%%%%%%%%%%%%%%%%%%%%%%%%%%%%%%%%%%%%%%%%%%%%%%%%%%%%%%%
%%%%%%%%%%%%%%%%%%%%%%%%%%%%%%%%%%%%%%%%%%%%%%%%%%%%%%%%%%%%%%%%%%%%%%%%
%%%%%%%%%%%%%%%%%%%%%%%%%%%%%%%%%%%%%%%%%%%%%%%%%%%%%%%%%%%%%%%%%%%%%%%%
\onecolumngrid
\begin{appendix}

\section{Derivation of Eq.~(\ref{eq:V3b})}
\label{app:E(R)}
As discussed in the main text around Eq.~(\ref{eq:deSitter}),
we take the metric outside the outer surface of the bubble shell, $\mathcal{H}_{\rm t}$, to be given by~(\ref{eq:deSitter}).
Since the hypersurface $\mathcal{H}_{\rm t}$ does not appear simple in this coordinate system,
we consider the following coordinate transformation:
\bea
\matrix{\dd t \\ \dd r \\ \dd\theta \\ \dd\phi} 
= \matrix{\e^{Ht} & \e^{Ht} v & 0 & 0 \\ 
                v &         1 & 0 & 0 \\ 
                0 &         0 & 1 & 0 \\ 
                0 &         0 & 0 & 1}
  \matrix{\dd t' \\ \dd r' \\ \dd\theta' \\ \dd\phi'}
\equiv \Lambda \matrix{\dd t' \\ \dd r' \\ \dd\theta' \\ \dd\phi'}.
\label{eq:coordinatestransformation} 
\eea
The primed coordinates are valid only in an infinitesimal region of $\mathcal{H}_{\rm t}$ around a given point on $\mathcal{H}_{\rm t}$.
Let us call this infinitesimal region $\delta\mathcal{H}_{\rm t}$, and
$v$ is constant within $\delta\mathcal{H}_{\rm t}$.
We would like $\delta\mathcal{H}_{\rm t}$ to be locally described as a surface on which $\dd r' = 0$.
Through~(\ref{eq:coordinatestransformation}), this corresponds to $v = \e^{Ht} \dd r / \dd t$.
Thus, $v$ can be interpreted as the ``proper speed'' of the expansion of the bubble's outer surface as measured in the unprimed coordinates under the metric~(\ref{eq:newmetric}).
In other words, the primed coordinates can be regarded as the instantaneous rest frame of a point on the outer surface.
Since $\Lambda$ is not invertible at $v=1$, 
we take $v$ in the range $0 \leq v < 1$,
which is also physical because the bubble expansion speed is near 1 but not exactly 1.

In~(\ref{eq:coordinatestransformation}), 
the $\Lambda^t_{~t'}$ component has been chosen such that $t'$ would be the conformal time, 
while the $\Lambda^r_{~r'}$ component has been chosen such that $\dd r = \dd r'$ on any spacelike 3d hypersurface with a constant $t'$.
The $\Lambda^r_{~t'}$ component simply defines $v$ as discussed above.
The $\Lambda^t_{~r'}$ component has been chosen such that the metric remains diagonal in the primed coordinates:
\bea
(g_{\mu'\nu'}) 
= \Lambda^\mathrm{\!T} 
  \matrix{-1 &        0 &            0 &                             0 \\ 
           0 & \e^{2Ht} &            0 &                             0 \\
           0 &        0 & \e^{2Ht} r^2 &                             0 \\
           0 &        0 &            0 & \e^{2Ht} r^2 \sin^{2\!}\theta
         }
  \Lambda
= \matrix{-\e^{2Ht}(1-v^2) &               0 &            0 &                             0 \\ 
                         0 & \e^{2Ht}(1-v^2) &            0 &                             0 \\
                         0 &               0 & \e^{2Ht} r^2 &                             0 \\
                         0 &               0 &            0 & \e^{2Ht} r^2 \sin^{2\!}\theta
         }.
\label{eq:newmetric} 
\eea

Now, since $\dd r' = 0$ on $\delta\mathcal{H}_{\rm t}$,
the 3d induced metric in $\delta\mathcal{H}_{\rm t}$ can be immediately read off from~(\ref{eq:newmetric}) by removing the second row and second column from it:
\bea
(\gamma_{\mu'\nu'}) 
= \matrix{-\e^{2Ht}(1-v^2) &            0 &                             0 \\ 
                         0 & \e^{2Ht} r^2 &                             0 \\
                         0 &            0 & \e^{2Ht} r^2 \sin^{2\!}\theta
         },
\label{eq:inducedmetric} 
\eea
where $\mu',\nu' = t', \theta', \phi'$.
Hence,
\bea
\sqrt{-\gamma'} = \e^{3Ht} \sqrt{1-v^2} \, r^2 \sin\theta 
\,.\label{eq:detgammaprime}
\eea
In the primed coordinates, the normal vector on $\delta\mathcal{H}_{\rm t}$ can also be easily found from~(\ref{eq:newmetric}):
\bea
(n_{\mu\prime}) 
= (0, -\e^{Ht} \sqrt{1-v^2}, 0, 0)
\,,
\eea
where the negative sign is due to our convention that this vector should point inward from $\mathcal{H}_{\rm t}$ into $\mathcal{V}_{\rm 4b}$.
We also need $v^\mu$, which is by definition the 4-velocity of the observer who has been outside of the bubble and at rest in the \emph{unprimed} coordinates. 
So, it is given in the \emph{primed} coordinates at $\delta\mathcal{H}_{\rm t}$ by
\bea
(v^{\mu\prime})
= \Lambda^{\!-1} \matrix{1 \\ 0 \\ 0 \\ 0}
= \frac{1}{\e^{Ht}(1-v^2)} \matrix{1\\ -v \\ 0 \\ 0} .
\eea
Hence, with our approximation $T^{\mu\nu} = -\Delta V g^{\mu\nu}$ on $\mathcal{H}_{\rm t}$, we have
\bea
n_\mu T^{\mu\nu} v_\nu
\>=\> -\Delta V \, n \!\cdot\! v
\>=\> -\Delta V \frac{v}{\sqrt{1-v^2}}
\,,\label{eq:n.T.v}
\eea
which, together with~(\ref{eq:detgammaprime}), gives
\bea
\dd^3x \sqrt{-\gamma} \; n_\mu T^{\mu\nu} v_\nu
\>=\> \dd t' \, \dd\theta' \, \dd\phi' \, \e^{3Ht} \sqrt{1-v^2} \, r^2\sin\theta \, \frac{-\Delta V \, v}{\sqrt{1-v^2}}
\>=\> -\Delta V \, \dd r \, \dd\Omega \; \e^{3Ht} r^2
\,,\label{final_integrand}
\eea
where in the last step we have used the relation $\dd r = v \, \dd t'$, which follows readily from~(\ref{eq:coordinatestransformation}) as $\dd r' = 0$ on $\delta\mathcal{H}_{\rm t}$.
Since the rightmost expression in~(\ref{final_integrand}) no longer has any reference to the primed coordinates or to $\delta\mathcal{H}_{\rm t}$, 
it can be directly integrated over the entire $\mathcal{H}_{\rm t}$ in the original unprimed coordinates.  
This concludes the derivation of Eq.~(\ref{eq:V3b}) of the main text.

%%%%%%%%%%%%%%%%%%%%%%%%%%%%%%%%%%%%%%%%%%%%%%%%%%%%%%%%%%%%%%%%%%%%%%%%
%%%%%%%%%%%%%%%%%%%%%%%%%%%%%%%%%%%%%%%%%%%%%%%%%%%%%%%%%%%%%%%%%%%%%%%%

\section{The benchmark model and its properties}
\label{app:model}
Here we describe the benchmark model to generate the results presented in the paper. 

\subsection{The model}
\label{app:lagrangian}
We introduce a complex scalar field $\Phi$ charged under a U(1) gauge symmetry whose gauge field and coupling are $A_\mu$ and $g$.
We work at 1-loop in dimensional regularization in $4-\epsilon$ dimensions with $\overline{\text{MS}}$ subtraction.
Our model is defined in this scheme by the following lagrangian:
\bea
\mathcal{L} 
=-({\rm D}_{\!\mu}\Phi)^* \, {\rm D}^\mu\Phi - \frac{\lambda}{4} |\Phi|^4 
 +\frac{1}{g^2} \biggl[ -\frac{1}{4} F^{\mu\nu}F_{\mu\nu} - \frac{1}{2\xi} (\partial_\mu A^\mu)^2 \biggr]
 +\mathcal{L}_{\rm ct}
\,,\label{eq:U(1)-lagrangian}
\eea
where ${\rm D}_{\!\mu}\Phi \equiv (\partial_\mu -\I A_\mu ) \Phi$, and $\mathcal{L}_{\rm ct}$ denotes the 1-loop counterterms:
\bea
\mathcal{L}_{\rm ct} 
= \frac{1}{16\pi^2\epsilon} 
   \biggl[ -2(3 - \xi)g^2 ({\rm D}_{\!\mu}\Phi)^* \, {\rm D}^\mu\Phi 
           -\bigl( 5\lambda^2 - 4\xi\lambda g^2 + 24 g^4 \bigr) \frac{|\Phi|^4}{4}
           +\frac{2}{3} \, \frac{1}{4} F^{\mu\nu}F_{\mu\nu}
   \biggr]
\,,\label{eq:counterterms}
\eea
where the $1/\epsilon$ poles above are only the UV poles, and the treatment of IR divergence is implicit.
The ghost terms are implicit as they do not involve $\Phi$ nor $A_\mu$. The coefficient of $|\Phi|^2$ has been set to be zero in~(\ref{eq:U(1)-lagrangian}) and (\ref{eq:counterterms}) for simplicity so that
a first-order phase transition may be realized by adjusting the gauge coupling alone.
Although the absence of $|\Phi|^2$ (or the smallness of the $\Phi$ mass compared to all other scales in the problem) may be argued as fine-tuned, it is nonetheless completely unambiguous in the specified scheme at 1-loop, which is good enough for a benchmark model.

We would first like to calculate the 1PI effective potential $V_\text{eff}(\varphi)$
for $\varphi \equiv \sqrt2 \, Z^{-1/2} \langle0\lvert{\rm Re}[\Phi]\rvert0\rangle$ at 1-loop, 
where $\sqrt{Z}$ is the field-strength renormalization of $\Phi$, normalized such that $Z=1$ at tree level.
We also take $\varphi$ to be positive.
There is no loss of generality in looking at only the real and positive part of $\Phi$ because the effective potential is invariant under
the global part of the U(1).
It is not invariant, however, under the gauged part of the U(1). 
Nevertheless, physical quantities we need for our PBH calculation such as $\Gamma_{\rm n}$ and $\Delta V$ are gauge invariant (although not \emph{manifestly} invariant),
so we take the Landau gauge ($\xi \to 0$) to facilitate our calculation. Then, neglecting 2-loop contributions and beyond, we find
\beq
V_\text{eff}(\varphi) 
= \frac{\tilde{\lambda}}{16} \varphi^4
+ \frac{1}{64\pi^2} \!\sum_{i=1}^3  g_i \bigl[ M^2_i(\varphi) \bigr]^{2} \biggl(\ln\frac{M^2_i(\varphi)}{\mu^2} - c_i \biggr)
\,,\label{eq:CW}
\eeq
where $\tilde{\lambda} \equiv \lambda Z^2$, $\mu$ denotes the $\overline{\text{MS}}$ scale, 
$(c_1, c_2, c_3) = (3/2, \, 3/2, \, 5/6)$, 
$(g_1, g_2, g_3) = (1,1,3)$, and
\bea
M^2_1(\varphi) = \frac{3\lambda\varphi^2}{4}
\,,\quad
M^2_2(\varphi) = \frac{\lambda\varphi^2}{4}
\,,\quad
M^2_3(\varphi) = g^2 \varphi^2
\,,\label{eq:m2}
\eea
representing the 1-loop contributions from ${\rm Re}\,\Phi$, ${\rm Im}\,\Phi$, and $A_\mu$, respectively.
The couplings and $Z$ are $\mu$ dependent, 
for which the RG equations can be read off from~(\ref{eq:counterterms}) as
\bea
\mu\frac{\dd \ln Z}{\dd\mu} = \frac{1}{16\pi^2} 2 \!\cdot\! 3g^2 
\,,\quad
\mu\frac{\dd\lambda}{\dd\mu} = \frac{1}{16\pi^2} \bigl( 5\lambda^2 - 12\lambda g^2 + 24g^4 \bigr)
\,,\quad
\mu\frac{\dd}{\dd\mu} \frac{1}{g^2} = -\frac{1}{16\pi^2} \frac{2}{3}
\label{eq:RGEs}
\eea
at 1-loop. These in particular imply that
\beq
\mu\frac{\dd\tilde{\lambda}}{\dd\mu} = \frac{1}{16\pi^2} \bigl( 5 \tilde{\lambda}^2 + 24g^4 \bigr) 
\,.\label{eq:lambda_tilde_RGE}
\eeq
These RG evolutions  
guarantee at the 1-loop level that the effective potential $V_\text{eff}(\varphi)$ is independent of $\mu$, which then permits us to set $\mu$ to any scale we like.
We especially like $\mu = \mu_*$, where $\mu_*$ is the scale at which $\tilde{\lambda}=0$, 
because that makes both $M_1(\varphi)$ and $M_2(\varphi)$ vanish, 
leaving us with only the $M_3$ contribution to $V_{\rm eff}$.
Such value of $\mu_*$ always exists according to (\ref{eq:lambda_tilde_RGE}) with $\xi = 0$, provided that $g \neq 0$.
In terms of $\mu_*$, we simply have
\bea
V_{\rm eff}(\varphi) &=& 
\frac{3}{64\pi^2}
\!\left( \ln\frac{g^2\varphi^2}{\mu_*^2} - \frac56 \right)\!
\varphi^4 \,.
\label{eq:Veff_zeroT}
\eea
This simple expression of $V_{\rm eff}(\varphi)$ shows that $V_{\rm eff}(\varphi)$ has no meta-stable minima but has a unique global minimum at $\varphi = v_\varphi$, where 
\bea
v_\varphi 
= \e^{1/6} \frac{\mu_*}{g}  
\,.
\eea
The point $\varphi=0$ is a local maximum, so there are no false vacua. 
All we have is the true vacuum at $\varphi = v_\varphi$.
The height of the local maximum with respect to the vacuum, $\Delta V$, is given by 
\bea
\Delta V \equiv V_{\rm eff}(0) - V_{\rm eff}(v_\varphi) = \frac{3 v_\varphi^4}{128\pi^2} 
\simeq \!\left(\! 0.26\,\frac{\mu_*}{g} \!\right)^{\!\!4}
\,.\label{eq:DeltaV}
\eea

Now, with a non-zero temperature, a meta-stable local minimum develops at $\varphi = 0$.
In the path-integral representation of the partition function, we integrate over $A_\mu$ and ${\rm Im}[\Phi]$, 
but not over ${\rm Re}[\Phi]$, to get the effective Hamiltonian $H_{\rm eff}[\phi, T]$ for $\phi \equiv {\rm Re}[\Phi]$.
This $\phi$ should be distinguished from $\varphi$.
The former is the microscopic label ``$i$'' on $E_i$ in $\sum_i \e^{-\beta E_i}$,  
while the latter is a macroscopic thermodynamic quantity obtained from averaging over $\phi$ with the weight $\propto \e^{-\beta E[\phi]}$. 
Repeating a calculation formally similar to that for $V_{\rm eff}(\varphi)$ on a periodic imaginary time in the $\xi = 0$ gauge with $\mu = \mu_*$, we find $V_{\rm eff}(\phi, T)$ inside $H_{\rm eff}[\phi, T]$ to be given by
\beq
V_{\rm eff}(\phi, T) 
= V_{\rm eff}(\phi) + \frac{3T^4}{2\pi^2} \, J_{\rm B\,} \!\!\left( \!\frac{g^2\phi^2}{T^2}\! \right)\!
\label{eq:Veff_T}
\eeq
with
\beq
J_{\rm B}(x^2) \equiv \int_0^{\infty} \!\!\! \dd y \, y^2 \ln\Bigl( 1 - \e^{-\sqrt{y^2 + x^2}} \Bigr)
\,.
\eeq
Since $J_{\rm B}(0) = -\pi^4/45 < 0$ and $J_{\rm B}(x^2)$ is a monotonically increasing function of $x^2$, 
a shallow local minimum develops at $x=0$ for $T>0$.
The depth of the local minimum is given by
\beq
V_{\rm eff}(0) - V_{\rm eff}(0,T) = -\frac{3T^4}{2\pi^2} J_{\rm B}(0) = \frac{\pi^2}{30} T^4
\,.\label{eq:depth}
\eeq

\subsection{The bubble nucleation rate}
\label{app:nucleation_rate}
To compute the bubble nucleation rate per unit volume, $\Gamma_{\rm n}$, 
there are two processes that can contribute to $\Gamma_{\rm n}$.
One is via thermal fluctuation, while the other is through quantum tunnelling.
Let us begin with the former. 
For simplicity we approximate that $\phi$ can be treated classically.
(This approximation will be retroactively justified by the largeness of $E[\phi_\C]/T$ defined below and plotted in Fig.~\ref{fig:S3overT}.)
Then, since the system before the phase transition is nearly in statistical equilibrium, 
$\Gamma_{\rm n}$ is proportional to the Boltzmann factor $\e^{-E[\phi_\C]/T}$, where $E[\phi]$ is the energy in a classical field  configuration $\phi$, and $\phi_\C$ is a specific configuration to be described shortly.
In calculating $E[\phi]$, 
we ignore loop corrections to the kinetic term $(\partial\phi)^2$ and all loop-induced terms containing derivatives of $\phi$ in the 1PI effective Lagrangian 
by assuming that they do not modify the physics provided by the tree-level kinetic term
unlike the loop corrections to the potential that qualitatively alter the physics expected from the tree-level potential.
Then, with $(\partial\phi)^2$, minimizing the energy to maximize $\e^{-E[\phi_\C]/T}$ leads us to considering only a static and spherically symmetric classical field configuration of $\phi$, while still keeping the dependence on $r$ to accommodate the boundary condition that $\phi$ should approach the false vacuum as $r \to \infty$ without making everywhere the false vacuum. 
For such $\phi$, then, $E[\phi]$ is given by
\beq
E[\phi] 
= 4\pi \!\int_0^\infty\!\!\!\! \dd r \> r^2 
  \biggl[ \frac12 \Bigl( \frac{\dd\phi}{\dd r} \Bigr)^{\!2} + V_{\eff}(\phi, T) - V_{\rm eff}(0,T) \biggr] 
\,.\label{eq:E_phi}
\eeq
Now, $\phi_\C(r)$ is so-called the \emph{critical bubble} solution, i.e., 
the lowest-energy static bubble that is unstable and would expand and fill the universe with the true vacuum if the value of $\phi$ at the center were slightly larger, 
or would collapse back to a point and leave the universe in the false vacuum if it were slightly smaller.
In other words, $\phi_\C(r)$ is the $\phi(r)$ that minimizes $E[\phi]$ and satisfies the boundary conditions $\phi(0) = \phi_0 \neq 0$ and $\phi(\infty) = 0$. 
Here, $\phi_0$ is not arbitrary but must be fixed by the condition that $\phi_\C(r)$ be smooth at the center, $\phi'_\C(0) = 0$, to avoid an infinite gradient energy density at the center.
We use the CosmoTransitions v.2.0.2 package\,\cite{Wainwright:2011kj} to compute $\phi_\C(r)$ and $E[\phi_\C]$ numerically. 
(In the package, and also commonly in the literature, $E[\phi]$ is called $S_3[\phi]$ because this is a mechanics problem in which $S_3$ is the ``action'' with $r$ and $\phi$ being the ``time'' and ``position'' of a particle.)
Setup parameters we used in the package are given by ${\rm alpha}=2$, ${\rm phi\_eps}=10^{-14}$--$10^{-8}$ for the SingleFieldInstanton command, and ${\rm xtol}=10^{-14}$--$10^{-8}$, ${\rm phitol}=10^{-17}$--$10^{-11}$, ${\rm rmax}=10^{15}$, ${\rm npoints}=50000$ for the findProfile command depending on temperatures (we require higher precision for lower temperatures).

The prefactor of the exponential in $\Gamma_{\rm n} \propto \e^{-E[\phi_\C]/T}$ is difficult to calculate.
Since the bubble nucleation is inherently a non-equilibrium process, the proportionality factor cannot be calculated using statistical mechanics alone as it must involve kinetic quantities like how frequently the interaction of a region of the false vacuum with its environment gives enough energy to the region for it to fluctuate into a critical bubble.
To remove extreme dimensionless quantities from the consideration, we assume that the U(1) gauge coupling $g$ in~(\ref{eq:U(1)-lagrangian}) is roughly $O(1)$.
(Recall that the other dimensionless coupling, $\lambda$, has been traded for a dimensionful parameter $\mu_*$ via the condition $\lambda(\mu_*) = 0$.)
We are then left with two dimensionful quantities $\mu_*$ and $T$.
Since we are interested in a supercooled phase, in which the depth of the false vacuum~(\ref{eq:depth}) is much smaller than the $\Delta V$ in~(\ref{eq:DeltaV}), we have $T \ll \mu_*$.

So, the question is which of the two scales governs the prefactor.
We would like to argue that it should be $T$.
In order for $\phi$ to escape from the false vacuum, it needs to ``collide'' with something in the bath and pick up enough energy to go over the potential barrier.
The Boltzmann factor $\e^{-E[\phi_\C]/T}$ accounts for the rarity for $\phi$ to pick up enough energy $E[\phi_\C]$ from any \emph{given} collision, but it does not account for the rate of how often collisions themselves occur.
The latter is encoded in the prefactor.
Before the system escapes from the false vacuum, $\phi$ is only fluctuating around $\phi=0$ where $V_{\rm eff}(\phi)$ is nearly flat. 
Thus, the dynamics of the system before the escape is nearly insensitive to $\mu_*$, which makes $T$ the only relevant dimensionful quantity in the problem.
Therefore, it seems reasonable to argue that the collision rate is governed by $T$, and hence the dimensionful part of the prefactor should be $T^4$.
The prefactor also contains the dimensionless normalization factor of the probabilities $\propto \e^{-E[\phi]/T}$.
But, again, since the system before the escape is nearly in statistical equilibrium at temperature $T$, the normalization factor is dominated by the low-energy states with $E \sim T$ fluctuating around $\phi=0$ so it cannot sensitively depend on $\mu_*$. The normalization factor, being dimensionless, should then be nearly constant.
The prefactor, overall, can have a mild logarithmic dependence on $\mu_*/T$ because from~(\ref{eq:Veff_T}) with~(\ref{eq:Veff_zeroT}) we see that a significant deviation from $O(1)$ in the value of $\phi/T$, when it happens, would be via $\ln(\mu_*/T) \gg 1$. 
We ignore such mild dependences as well as any possible numerical factors like $\pi$.
Therefore, for our estimates, we use the expression
\bea
\Gamma_{\rm n}(T) = T^4 \, \e^{-E[\phi_\C]/T}
\,.\label{Gamma_n_approx}
\eea

Next, we define the nucleation temperature $T_{\rm n}$ via $\Gamma_{\rm n}(T_{\rm n}) = H^4$, 
where $H^2 = 8\pi\Delta V / (3M_{\rm Pl}^2)$, 
neglecting both the dip of $\pi^2 T^4/30$ and the energy density in radiation compared to $\Delta V$, 
which is a good approximation in a supercooled phase.
This definition combined with~(\ref{Gamma_n_approx}) can then be inverted to express the required value of $E[\phi_\C]/T$ for the bubble nucleation rate to match the expansion rate at $T=T_{\rm n}$:
\bea
\frac{E[\phi_\C]}{T} \biggr|_{T = T_{\rm n}} 
=\> \ln \!\left[ \!\left( \frac{3M_{\rm Pl}^2 T_{\rm n}^2}{8\pi\Delta V} \right)^{\!\!2} \right]\! 
\>\simeq\>
6.5 + 4\ln\!\left[ \frac{gT_{\rm n}}{\mu_*} \frac{gM_{\rm Pl}}{\mu_*} \right]\!
%\,,\label{Tn_condition}
\eea
where $M_{\rm Pl} = 1.22 \times 10^{19}\,\GeV$, 
and we have used~(\ref{eq:DeltaV}) in the last step.

\begin{figure}[t] 
\begin{center}
\includegraphics[width=0.45\textwidth]{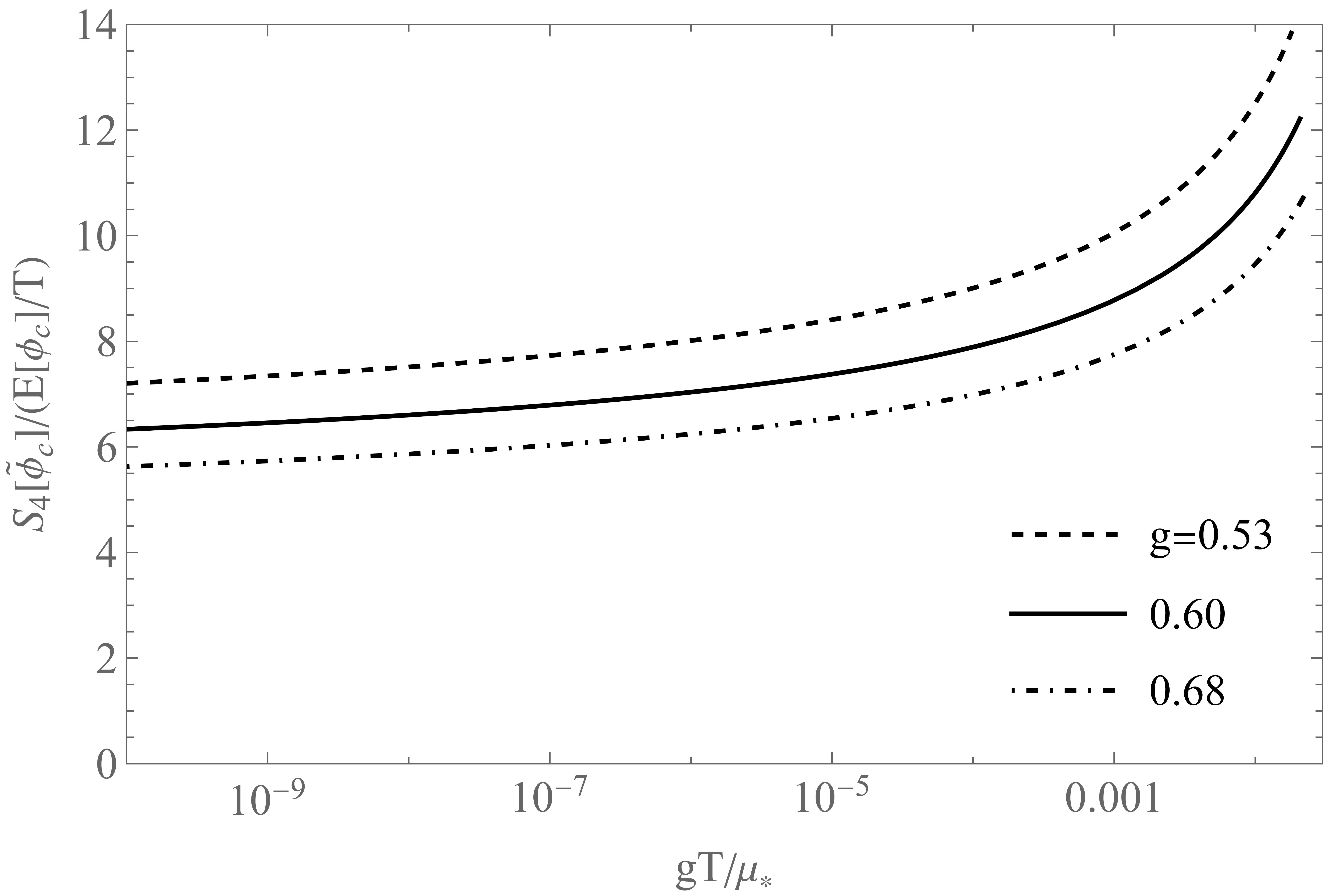} 
\end{center}
\caption{Comparison between $E[\phi_\C]/T$ and $S_4[\tilde{\phi}_\C]$ as a function of $gT/\mu_*$ for different values of $g$. $S_4[\tilde{\phi}_\C]$ is always greater than $E[\phi_\C]/T$, so the quantum tunnelling contribution to $\Gamma_{\rm n}$ is always exponentially smaller than the thermal one.}
\label{fig:S4vsS3}
\end{figure}

It is clear from dimensional analysis that $E[\phi_\C]/T$ depends on $T$ and $\mu_*$ only through the combination $T/\mu_*$.
Moreover, $\mu_*$ always comes in the combination $\mu_*/g$.
In Fig.~\ref{fig:S3overT}, the black curves show $E[\phi_\C]/T$ as a function of $gT / \mu_*$ for different values of $g$,
while the straight lines show the value of the right-hand side of~(\ref{Tn_condition}) as a function of $gT/\mu_*$ for different values of $\mu_*/g$.
For each value of $g$, 
as we lower the temperature and follow its black curve from top-right to lower-left, 
when the curve meets the line of given $\mu_*/g$, that intersection gives $T_{\rm n}$ for that $\mu_*/g$.
Since the right-hand side of~(\ref{Tn_condition}) says that the vertical offset of the line increases as $\mu_*/g$ decreases, the curve is guaranteed to intersect with the line for any given $g$ if we choose $\mu_*/g$ to be sufficiently low.
When $\mu_*/g$ gets too high, no intersections exist and the bubble nucleation rate would never be large enough to catch up with the expansion rate of the universe.
This feature was also pointed out in Ref.\,\cite{Ellis:2018mja, Lewicki:2021xku}.

Let us now discuss the quantum tunneling contribution to $\Gamma_{\rm n}$. 
The exponential part of that contribution is given by $\e^{-S_4[\tilde{\phi}_\C]}$, where $S_4[\phi]$ is given by~\eqref{eq:E_phi} with the replacement $4\pi r^2 \to 2\pi^2 r^3$ and $\tilde{\phi}_\C$ is like $\phi_\C$ except that it is from minimizing $S_4$ instead.
We take the prefactor of the exponential to be $T^4$ again by the same argument.
In Fig.~\ref{fig:S4vsS3} right, we see that $S_4$ is always greater than $E/T$ so the quantum tunneling contribution is negligible.

To use $\Gamma_{\rm n}$ for the estimation of the PBH abundance described in~\ref{sec:PBHabundance}, 
we need $\Gamma_{\rm n}$ as a function of $t$ rather than $T$. 
To relate them we use $T(t) = T_{\rm n} \, \e^{-H(t-t_{\rm n})}$. 
This should be a good approximation as long as the universe is dominated by the vacuum energy, which is the case during the bubble formation period except near the very end of the phase transition when the spacetime is no longer approximately de Sitter.

Once $\Gamma_{\rm n}(t)$ is thus obtained, we calculate $P_{\rm f}(t)$, $P_\PBH(t)$, and $n_\PBH$ by~\eqref{eq:P_f}, \eqref{eq:P_PBH}, and \eqref{eq:nPBH}, respectively. 
In Fig.~\ref{fig:benchmark_fixed_mustar}, $P_{\rm f}(t)$, $P_\PBH(t)$, and the \emph{integrand} of $n_\PBH$ are plotted for the benchmark values of $\mu_* = 10^6\,\GeV$ and three different values of $g$, all with $\kappa=1$ and $\Delta\Omega/4\pi=10^{-3}$.

\end{appendix}
%%%%%%%%%%%%%%%%%%%%%%%%%%%%%%%%%%%%%%%%%%%%%%%%%%%%%%%%%%%%%%%%%%%%%%%%
%%%%%%%%%%%%%%%%%%%%%%%%%%%%%%%%%%%%%%%%%%%%%%%%%%%%%%%%%%%%%%%%%%%%%%%%
%%%%%%%%%%%%%%%%%%%%%%%%%%%%%%%%%%%%%%%%%%%%%%%%%%%%%%%%%%%%%%%%%%%%%%%%
%%%%%%%%%%%%%%%%%%%%%%%%%%%%%%%%%%%%%%%%%%%%%%%%%%%%%%%%%%%%%%%%%%%%%%%%

%\bibliographystyle{utphys}
%\bibliography{ref}

\providecommand{\href}[2]{#2}\begingroup\raggedright\endgroup

\end{document}